\pdfoutput=1
\documentclass[preprint,12pt,authoryear]{elsarticle}

\usepackage[T1]{fontenc}
\usepackage[utf8]{inputenc}
\usepackage{lmodern}
\usepackage{amsmath,amssymb}
\usepackage{graphicx}
\usepackage{booktabs}
\usepackage{tabularx}
\usepackage{array}
\usepackage{multirow}
\usepackage{makecell}
\usepackage[table]{xcolor}
\usepackage{caption}
\usepackage{subcaption}
\usepackage{seqsplit}
\usepackage{xspace}
\usepackage{url}
\usepackage[breakable]{tcolorbox}
\usepackage[hidelinks]{hyperref}

\graphicspath{{figures/}}

\providecommand{\ie}{i.e.,\xspace}
\providecommand{\eg}{e.g.,\xspace}

\providecommand{\icode}[1]{\texttt{#1}}
\tcbset{markedbreak/.style={breakable}}

\makeatletter
\def\ps@pprintTitle{%
  \let\@oddhead\@empty
  \let\@evenhead\@empty
  \def\@oddfoot{\reset@font\hfil\thepage\hfil}%
  \let\@evenfoot\@oddfoot}
\makeatother

\begin{document}

\begin{frontmatter}

\title{Characterizing and Predicting the Longest CI Builds in Machine Learning Projects}

\author[otago]{Yonghui (Andie) Huang\corref{cor1}}
\ead{andie.huang@postgrad.otago.ac.nz}
\author[otago]{Daniel Alencar da Costa}
\ead{danielcalencar@otago.ac.nz}
\author[otago]{Grant Dick}
\ead{grant.dick@otago.ac.nz}
\author[zewail]{Mohamed Sami Rakha}
\ead{rakha@zewailcity.edu.eg}

\cortext[cor1]{Corresponding author.}

\affiliation[otago]{organization={School of Computing, University of Otago},
            addressline={133 Union Street East},
            city={Dunedin},
            postcode={9016},
            country={New Zealand}}

\affiliation[zewail]{organization={Zewail City of Science and Technology},
            city={6th of October City},
            state={Giza},
            country={Egypt}}

\begin{abstract}
Continuous integration (CI) builds in machine learning (ML) projects tend to run longer than in other projects, yet little is known about the longest builds, which consume most CI time.
This paper characterizes the longest builds in ML projects, analyzes how CI workflow properties are associated with them, and evaluates whether they can be predicted before execution.
We study 269,930 GitHub Actions builds from 107 open-source ML projects.
The top 18\% longest builds account for 80.30\% of total build time and fail in 39.09\% of cases.
Build duration differs significantly across ML project types. The framework and infrastructure ML projects have the longest upper tail and produce 98\% of the longest builds, whereas projects that use existing ML libraries have the lowest median build duration.
The longest builds are most strongly associated with the recent history of the same workflow, as the duration of its last build and the median duration of the project's last 20 builds show large effect sizes.
The longest builds are more consistently associated with CI workflow structure than with configuration.
The number of jobs, control flow complexity, coupling complexity, and maximum steps per job show medium effect sizes, whereas caching, matrix settings, and Docker builds show small or negligible effects.
Framework-specific workflows have the largest percentage of builds among long builds. Platform-specific workflows produce the largest number of the longest builds.
Finally, 47 features describing the build, the recent builds of the same workflow, its workflow structure, and the ML usage depth and domain of the project outperform a traditional build-duration feature baseline with all four classifiers.
When builds are evaluated in chronological order, the best model (XGBoost) reaches an F1-score of 0.775, compared with 0.682 for the baseline, and an AUC-ROC of 0.955.
These results show that build history and CI workflow structure help identify the long-running builds in ML projects before they run.
\end{abstract}


\end{frontmatter}

\section{Introduction}
\label{sec:longbuilds:intro}
Software products increasingly use machine learning (ML) in applications such as search ranking, recommendation systems, and fraud detection \citep{liu2009learning,adomavicius2005next,ali2022financial}. Once ML functionality is deployed, maintaining it becomes a recurring engineering task. Incoming data may change, the deployed model may become outdated as the business grows, and the surrounding software stack may evolve \citep{sculley2015hidden,amershi2019software,kreuzberger2023machine}. \citet{sculley2015hidden} show that ML systems often depend on data formats, feature definitions, libraries, and runtime versions. As a result, retraining and redeployment are not only model updates but also require teams to keep data preparation, dependencies, and execution environments consistent. At the same time, continuous integration (CI) has become a standard way to keep traditional software in a releasable state \citep{shahin2017continuous}. Industrial Machine Learning Operations (MLOps) guidelines extend this idea and recommend that~the ML project development process should adopt automated CI/CD pipelines in order to make the whole project delivery repeatable and observable \citep{kreuzberger2023machine}.

From a software engineering perspective, ML projects are more complex because of data, model, and infrastructure dependencies and iterative workflows~\citep{sculley2015hidden,amershi2019software}.
In this paper, we use \textit{ML projects} to refer to software projects whose repositories contain code for building, using, or supporting ML functionality.
This definition covers projects that provide ML infrastructure and projects that apply ML to specific tasks.
In both cases, the software depends not only on the source code but also on the data used by the project, the model training process, the execution environment, and the services connected to it~\citep{sculley2015hidden,amershi2019software}.
As a result, even a routine change to data, dependencies, or infrastructure can require teams to rebuild, test, and revise several parts of the project.
These characteristics affect how teams design CI/CD pipelines and how they experience build and deployment activities in ML projects.
Recent research has started to examine how ML projects use CI/CD in practice. For example, \citet{rzig2022characterizing} characterize CI adoption in ML projects by studying which CI services they use more frequently and how often the developers modify CI configurations. \citet{houerbi2024empiricalarxiv} analyze how CI/CD configurations change in ML projects, identify common change categories and 14 co-change patterns with machine learning related code, and report frequent changes to CI/CD rules together with recurring configuration problems, such as adding dependencies directly in CI workflow files and keeping outdated settings in the configuration. \citet{shahin2017continuous} identify long build and testing times and unreliable deployment pipelines as common challenges in adopting CI/CD. Taken together, these studies suggest that CI/CD workflows in ML projects are frequently revised as project dependencies, build steps, and configuration rules change. Such changes may affect how builds are executed and, in turn, may influence build duration in ML projects \citep{shahin2017continuous,rzig2022characterizing,houerbi2024empiricalarxiv}.

Among build-duration issues, the longest builds are particularly important because they can consume disproportionate CI resources and delay developer feedback.
As a result, understanding which builds become the longest can help teams identify where CI time is concentrated and manage CI resources more effectively.

However, no prior work systematically characterizes these longest builds or predicts their occurrence in ML projects to support better resource management for CI execution.
There is also emerging evidence that CI/CD pipelines in ML projects often have longer build durations than pipelines in non-ML projects~\citep{bernardo2024machine}.
\citet{bernardo2024machine} compared CI build durations in ML projects with matched non-ML projects and found that builds in ML projects tend to run longer, while medium-sized ML projects exhibit lower test coverage than non-ML projects.
They also report that developers see build duration and build stability as specific obstacles when applying CI to ML projects~\citep{bernardo2025continuous}.
However, despite their practical importance, the longest builds in ML projects remain poorly understood.
Prior work has not systematically examined which builds become the longest, what project and CI workflow characteristics are associated with them, or whether long duration builds can be predicted before execution.
This gap limits our ability to understand where CI time is concentrated and how teams might anticipate costly CI runs.
As a result, practitioners still lack empirical guidance on how to identify and anticipate the longest builds in ML projects so that CI resources can be allocated more effectively.

This lack of focused evidence is problematic for both research and practice. MLOps guidance recommends automated CI pipelines to help teams test changes, monitor model behavior, and retrain models when data or system conditions change \citep{kreuzberger2023machine}. Such practices are important because ML service reliability depends not only on source code, but also on data and model behavior. When CI builds take a long time to finish, teams receive feedback later, which can slow experimentation, delay defect detection, and increase infrastructure cost \citep{shahin2017continuous,bernardo2025continuous,kreuzberger2023machine}. This issue is especially relevant to ML projects, where CI builds may include work beyond compiling code and running tests. Build duration may also be affected by data handling steps, model training stages, and the configuration of specialized hardware \citep{amershi2019software,bernardo2024machine}. A systematic investigation of the longest builds is therefore needed to identify which CI workflow and ML project characteristics are associated with long build duration, and to help teams anticipate builds that may consume large CI resources before they occur.

To address this gap, we conduct an empirical study of CI builds from 107 open-source ML projects that use GitHub Actions.
In this paper, \textit{longest builds} refers to the builds located in the upper tail of the build duration distribution among our studied dataset.
Following the Pareto principle~\citep{fenton2014software}, we define them as the smallest top percentile of builds whose cumulative duration reaches 80\% of the total build time.
In our dataset, \textbf{the longest builds are the top 18\% of builds}, \ie~builds that take at least 39.20 minutes. Section~\ref{approach:longbuilds:RQ1} describes how this threshold is derived.
Specifically, we aim to characterize the longest builds in ML projects, examine how CI workflow properties (e.g., the organization of jobs, steps, and dependencies in CI workflows) and ML project attributes are associated with the occurrence of the longest builds, and evaluate whether these characteristics can be used to predict such builds before execution.
We base our study on prior work comparing CI practices in ML and non-ML projects and examining CI/CD evolution in ML projects~\citep{rzig2022characterizing,bernardo2024machine,houerbi2024empiricalarxiv}. Our sample is drawn from the curated project lists used in these studies, which in turn originate from~\citet{gonzalez2020state}, and is further filtered to retain repositories with active GitHub Actions workflows and sufficient CI run history. Specifically, we investigate the following research questions:

\begin{itemize}
    \item \textbf{RQ1:} How can the longest builds in ML projects be characterized?

    \item \textbf{RQ2:}
    How are CI workflow characteristics associated with the longest build durations in ML projects?

    \item \textbf{RQ3:} Can the longest builds in ML projects be predicted accurately before execution?
\end{itemize}

Through statistical analysis on our dataset of 269,930 builds from 107 ML projects, we find that total build time is highly concentrated in a small fraction of builds. In particular, the top 18\% longest builds account for about 80\% of total build time and exhibit failure rates of about 39\%.

We therefore focus on the top 18\% longest builds, because they consume most of the CI time.
We also find that the longest build durations are more consistently associated with how jobs depend on each other within a CI workflow than with execution environment factors, such as the operating system or CI runner configuration.
Finally, we try to predict which builds will be among the top 18\% longest builds before they run.
Our features describe the recent builds and the structure of the CI workflow, perform better than the build features used in prior work~\citep{ghaleb2019empirical} on predicting the top 18\% longest builds. The best model reaches an F1 of 0.775 and an AUC-ROC of 0.955.

This work contributes to the literature on software engineering for ML and on MLOps in three ways:

\begin{itemize}
    \item We provide a systematic characterization of build duration in ML projects. We first analyze the distribution of build duration across the studied projects, and then focus on the top 18\% longest builds. This analysis shows that these longest builds account for most of the total CI time and appear in most of the studied projects in our dataset. We also classify the projects by how they use ML and by their application area, and then analyze how build duration differs across these project categories.

    \item We analyze how CI jobs are organized within CI workflows and identify CI workflow characteristics associated with the longest builds. The results show that the longest builds are more common when CI workflows contain more dependencies among jobs and longer chains of dependent jobs. In contrast, execution environment factors, such as the operating system and CI runner configuration, show weaker associations.

    \item We propose 47 features to predict the longest builds in ML projects before they run. They describe the current build, recent builds of the same workflow, its job structure, and the project's ML usage depth and domain. These features outperform the build features used in prior work on predicting build duration in general software projects~\citep{ghaleb2019empirical}.
\end{itemize}

These contributions build on existing studies of CI in ML projects~\citep{rzig2022characterizing,bernardo2024machine,houerbi2024empiricalarxiv,bernardo2025continuous}.
While prior work has examined CI usage, configuration changes, and differences between ML and non-ML projects, this paper focuses on build duration and on the small group of builds that consume most CI time.
It also complements broader work on the maintenance challenges and development processes of ML projects~\citep{sculley2015hidden,amershi2019software} by showing how these challenges are reflected in CI build duration.

The remainder of this paper is organized as follows. Section~\ref{sec:longbuilds:methodology} presents the study design, including project selection, data collection, and feature engineering. Section~\ref{sec:longbuilds:rqs} presents the motivation, approach, and findings for each research question. Section~\ref{sec:longbuilds:discussion} discusses the implications of our findings for developers and researchers. Section~\ref{sec:longbuilds:threats} outlines threats to validity, Section~\ref{sec:longbuilds:related} positions this study against prior work on long CI builds, ML CI workflows, and CI cost reduction, and Section~\ref{sec:longbuilds:conclusion} concludes and suggests directions for future work.

\section{Study Design}
\label{sec:longbuilds:methodology}
\subsection{Studied Projects}
Our dataset construction follows the ML project selection pipeline proposed by \citet{gonzalez2020state}, refined by \citet{rzig2022characterizing}, and reused by \citet{bernardo2024machine}. We adapt this pipeline with filters for GitHub Actions, informed by prior studies of GitHub Actions and CI adoption on GitHub~\citep{vasilescu2015quality,bernardo2018studying,decan2022use}.

\subsubsection{Starting point: ML projects on GitHub}
We begin with the ML project category defined by \citet{gonzalez2020state} and reused by \citet{rzig2022characterizing} and \citet{bernardo2024machine}. This category includes ML frameworks and libraries (\eg~\texttt{scikit-learn}) and ML applications (\eg~\texttt{faceswap}).
\citet{rzig2022characterizing} revised the original list by manually reading project descriptions on the main GitHub pages and the websites linked from those pages. They removed tutorials, study guides, toy repositories, and repositories that did not actually use ML. The curated dataset contains 3{,}981 ML projects and 4{,}026 non-ML projects. We use the 3{,}981 curated ML projects as our starting set and apply the subsequent filtering steps described below.

\subsubsection{Step 1: Activity and default branch filtering}
First, we restrict attention to projects that were still active in 2025, so that the dataset reflects recent development and CI practices rather than abandoned repositories. Using the GitHub REST API\footnote{\url{https://docs.github.com/en/rest}}, we query the ML repositories and retain those whose default branch was updated at least once between January 1 and July 11, 2025. We also require the default branch name to be either \textbf{main} or \textbf{master}. To keep branch selection comparable across repositories and consistent with prior criteria~\citep{bernardo2024machine}, we discard repositories whose default branch uses other names, such as dev, development, or draft. This step yields 2{,}705 repositories.

\subsubsection{Step 2: Presence of GitHub Actions workflows}
Second, we remove projects that do not use GitHub Actions, because this study focuses on CI builds executed on GitHub Actions. A repository is considered to use GitHub Actions if it contains at least one workflow configuration file under the \path{.github/workflows/} directory.
In this paper, a \textit{workflow} refers to a YAML configuration file that specifies what should run after a repository event, such as a push or pull request. A workflow is made up of \textit{jobs}, and each job contains an ordered list of related \textit{steps}. A \textit{build} refers to one execution of a workflow in the CI system~\citep{decan2022use,wessel2023github}.
We exclude repositories that do not have any such YAML file. After this filter, 1{,}009 ML repositories remain.

\subsubsection{Step 3: Identification of CI workflows}
GitHub Actions workflows can be used for many purposes beyond CI, such as documentation generation, issue triage, or community management~\citep{decan2022use}. For our study, we focus on workflows that implement CI activities. We label a workflow as a CI workflow only when it satisfies both of the following criteria. First, the workflow must be triggered by code changes. We require the \verb|on| field of the workflow file to include either \verb|push| or \verb|pull_request|. Second, at least one job name or step name in the workflow must contain CI terminology. We search for the following terms after converting names to lower case: \texttt{\seqsplit{"continuous\_integration"}}, \texttt{"continuous-integration"}, \texttt{"continuousintegration"}, and \texttt{"test"}. This criterion follows prior work that treats test execution as a core CI activity~\citep{bernardo2024machine}. Projects that do not contain any CI workflow are excluded. This step results in 454 ML repositories with at least one CI workflow built on GitHub Actions.

\subsubsection{Step 4: Minimum number of CI runs}
To ensure each project has enough CI activity for analysis, we apply a minimum threshold for the number of workflow runs. Prior work on CI outcomes and delivery time has shown that small numbers of builds can lead to unstable estimates and noisy trends~\citep{vasilescu2015quality,bernardo2018studying}. Following \citet{bernardo2024machine}, we retain only projects whose GitHub Actions CI workflows have recorded at least 100 runs in total. Runs include executions triggered by either \verb|push| or
\verb|pull_request| events. Projects that do not meet this threshold are removed from the dataset. After this filter, 120 ML projects remain.

\subsubsection{Step 5: Minimum CI history length}
Finally, we require that CI workflows have at least six months of history, so that the observed build durations are based on workflows with sustained use rather than workflows that were newly introduced or used only for a short period. Inspired by previous studies~\citep{vasilescu2015quality,bernardo2018studying}, we enforce a minimum history of six months. For each project, we compute the time span between the first and last recorded run of its CI workflows. Projects with less than six months between these two timestamps are discarded. This last step yields the final dataset of 107 ML projects, each with at least one active GitHub Actions CI workflow, at least one hundred CI runs, and at least six months of CI history.

This selection procedure provides a dataset that is consistent with prior work on ML projects and CI on GitHub, while focusing on ML repositories that use GitHub Actions for CI.

\subsection{Data Collection}
\label{subsec:longbuilds:data_collection}
After selecting 107 ML projects, we systematically collect their historical build and workflow data from GitHub Actions via the GitHub REST API up to July 11, 2025.
For API endpoints that return paginated results, we follow the pagination links until no further page is available.
For each project, we retrieve all CI workflows and their historical builds. Each CI workflow is defined by a YAML configuration file. We collect these files because the duration of a CI workflow run depends not only on the execution time of its individual jobs, but also on how these jobs are organized. Jobs that have no dependency between them can run at the same time, whereas jobs connected through \texttt{needs} must wait for their prerequisite jobs to finish~\citep{lebeuf2018understanding,github_actions_jobs,github_actions_workflow_syntax}. A longer chain of dependent jobs can therefore increase the time before the CI workflow completes, while more independent jobs can increase available parallelism. This issue is especially relevant for ML projects because their CI workflows may connect data preparation, environment setup, model training, evaluation, packaging, and testing~\citep{houerbi2024empiricalarxiv}. As a result, the organization of jobs in a CI workflow can be one of the possible causes of long build duration and is directly related to RQ2 and RQ3. Therefore, the YAML files are needed to reconstruct CI workflow DAGs and compute CI workflow structure features.

\paragraph{Step 1: Project and workflow collection}
For each repository, we first query the GitHub REST API to obtain basic repository metadata (\eg~default branch, description, and topics) and to list all workflow configuration files under the \texttt{.github/workflows} directory.
We use the workflow endpoint\footnotemark\footnotetext{\texttt{/repos/\{owner\}/\{repo\}/actions/workflows}} to obtain workflow identifiers, names, states, and file paths.
We then use the workflow file paths to download the corresponding YAML configuration files for later construction of CI workflow DAGs.
In total, we collect 289 CI workflows and their YAML configuration files from the 107 ML projects.

\paragraph{Step 2: Collection of workflow runs and jobs}
We then collect execution data for every workflow by retrieving its historical workflow runs\footnote{\texttt{/repos/\{owner\}/\{repo\}/actions/workflows/\{id\}/runs}}.
For each run, we collect the run identifier, status (\eg~\textit{``queued''}, \textit{``in\_progress''}, \textit{``completed''}), conclusion (\eg~\textit{``success''}, \textit{``failure''}), triggering event, timestamps, and associated commit SHA.
We also retrieve the jobs belonging to each workflow run\footnote{\texttt{/repos/\{owner\}/\{repo\}/actions/runs/\{run\_id\}/jobs}} and record their names, status, conclusion, and timestamps.
Based on the recorded timestamps, we calculate the execution duration of each completed run as the time interval between the run start time (\textit{run\_started\_at}) and the completion time (\textit{updated\_at}).
For runs triggered by pull requests, we additionally query the Pull Request API\footnotemark\footnotetext{\texttt{/repos/\{owner\}/\{repo\}/pulls/\{pull\_number\}}} to obtain code churn information (\eg~the number of modified files and changed lines in the pull request). In total, we collect 287{,}541 run records from 289 workflows across 107 ML projects. Because some runs were retrieved more than once, we keep one record per run identifier, which removes 4{,}277 duplicate records and leaves 283{,}264 unique runs.

\paragraph{Step 3: Exclusion of re-run builds}
GitHub Actions allows developers to re-run a completed workflow run, for example after a failure caused by a flaky test~\citep{github_actions_rerun}. A re-run does not create a new run. Instead, it keeps the same run identifier, increments the attempt number (\textit{run\_attempt}), and replaces the run start time and completion time with those of the latest attempt, while the creation time (\textit{created\_at}) still refers to the first attempt.
The duration of such a run therefore does not describe one execution of the workflow. The interval from \textit{run\_started\_at} to \textit{updated\_at} covers only the latest attempt and ignores the earlier ones, whereas the interval from \textit{created\_at} to \textit{updated\_at} also includes the earlier attempts and the time that passed before the developer triggered the re-run, which can be hours or days.
Because our analyses compare the durations of individual builds, we exclude all runs with \textit{run\_attempt} $> 1$.
This step removes 13{,}334 runs (4.7\%) and leaves 269{,}930 builds from 289 workflows across 107 ML projects for our analysis.
For every remaining run, the run start time coincides with the run creation time, so its duration measures a single execution of the workflow.

\begin{figure}
    \centering
    \includegraphics[width=\linewidth]{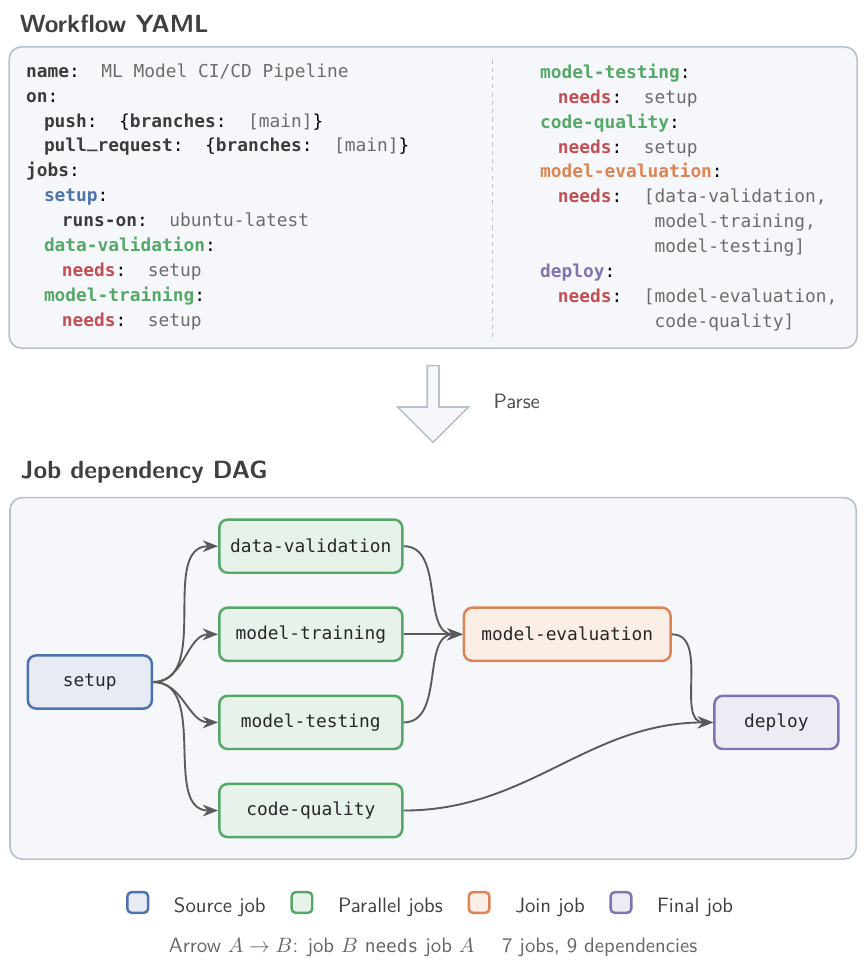}
    \caption{An example of DAG construction from workflow YAML.}
    \label{fig:dag-example}
\end{figure}

\paragraph{Step 4: Workflow DAG construction}
The purpose of this step is to transform each workflow YAML file into a measurable representation of job organization. A DAG is suitable for this purpose because it represents which jobs must finish before other jobs can start. These dependencies constrain which jobs can run in parallel~\citep{yin2009directed}. A graph representation makes these execution constraints explicit and allows us to compute workflow measures such as control flow complexity and coupling complexity.
This is especially relevant in ML projects, where prior work has shown that workflows are often more complex than in non-ML projects~\citep{bernardo2024machine}. We retrieve the YAML configuration file associated with each workflow using the GitHub contents API\footnote{\texttt{/repos/\{owner\}/\{repo\}/contents/\{workflow\_path\}}}. We then parse the YAML using the \texttt{PyYAML} library to extract the fields that describe workflow behavior, including trigger events (\texttt{on}), jobs (\texttt{jobs}), steps within each job, and dependencies among jobs declared through the \texttt{needs} attribute.
Using this parsed information, we construct a Directed Acyclic Graph (DAG) with the \texttt{NetworkX}\footnote{\url{https://networkx.org/}} library. Figure~\ref{fig:dag-example} shows how we transform a YAML workflow configuration into a DAG.

Each job identifier from the YAML \texttt{jobs} dictionary becomes a node in the DAG. If a job declares dependencies through \texttt{needs}, we normalize the value to a list and add one directed edge from each prerequisite job to the dependent job. Jobs without \texttt{needs} are included as nodes without incoming dependency edges. This process transforms the dependencies declared in workflow YAML into an explicit graph structure where nodes represent jobs and edges represent execution ordering constraints.
From each DAG, we compute workflow complexity measures that are used later in our feature engineering.
The graph metrics are joined with information about the build run, the CI workflow, and the project, which together form the set of 47 features used in our analyses and prediction models. In total, we build 289 DAGs, each associated with an active CI workflow from the 107 ML projects. These graphs provide the basis for the CI workflow complexity features used in RQ2 and RQ3.

\subsection{Feature Engineering}
\label{subsec:longbuilds:feature_engineering}
To predict the longest builds in ML projects, we construct a set of 47 features that capture information about the build run, the CI workflow, and the project.
The response variable is binary.
For each build, the label is 1 if the build falls within the longest 18\% of all builds, \ie~the smallest top percentile of builds that accounts for 80\% of the total build time, and 0 otherwise.
We explain how this threshold is derived in the approach of RQ1 (Section~\ref{approach:longbuilds:RQ1}).
All features are constrained to information that is available at build submission time.
We group these engineered features into seven families as summarized in Table~\ref{tab:feature_families}.
The project-level features that use manual labels reuse the final consensus labels produced in the project classification process for RQ1.
 The paragraphs below describe how the features are computed.

\subsubsection{Run-Level}
\paragraph*{(I) Code churn factors}
This family captures the size and location of the code change that triggers a CI run.
We include these features because prior studies on CI build duration and build outcome prediction have shown that the amount of changed code is an important signal for build cost and build failure risk~\citep{ghaleb2019empirical,jin2020cost}.
Larger changes may require more compilation, dependency resolution, and test execution, while changes to source, test, or build files may activate different parts of the CI workflow.
For each workflow run that is triggered by a pull request, we query the corresponding pull request of each build via the GitHub REST API.
The \emph{Number of commits}, \emph{Added code lines}, \emph{Deleted code lines}, and \emph{Total lines changed} are read directly from the pull request metadata (\texttt{commits}, \texttt{additions}, \texttt{deletions}, and \texttt{additions+deletions}, respectively).
We also use the changed file list of the pull request to count \emph{Modified source files}, \emph{Modified test files}, and \emph{Modified build files}.

\paragraph*{(II) Build history factors}
This family describes the earlier builds of the same project before a target build starts.
We include these features because prior CI studies show that recent build history is useful for predicting future build outcomes~\citep{rausch2017empirical,chen2020buildfast}.
For predicting long builds, the duration of earlier builds is also relevant. Prior work shows that the duration of a CI build is likely to be similar to the duration of earlier builds~\citep{ghaleb2022studying}.
We group the earlier builds of a target build in four ways, by the same CI workflow, the same CI workflow and branch, the same CI workflow and event, and the same project.
For a target build, we use only builds that finished \emph{before} the target build was created to avoid temporal leakage.
\emph{Last build duration} is the duration of the most recently finished earlier build.
\emph{Median duration of last 20 builds} is the median duration of the 20 most recently finished builds of the same project. For the earliest builds of a project, fewer than 20 builds have finished before the target build. In this case, we compute the median over all builds of the project that finished before the target build.
\emph{Duration variability of last 5 builds} is the standard deviation of the logarithm of build duration over the five most recently finished builds of the same CI workflow and branch. A standard deviation needs at least two builds. When fewer than two builds of the same CI workflow and branch have finished before the target build, we set this feature to 0.
\emph{Failure rate of last 10 builds} is the fraction of the ten most recently finished builds of the same CI workflow and event that ended in failure.
\emph{Builds since last failure} counts the successful builds of the same CI workflow that finished after its most recent failed build and before the target build was created.
We include these two failure features because prior work shows that build breakages are related to build duration. Actions to fix build breakages, such as retrying failed commands, are associated with longer builds~\citep{ghaleb2019empirical,ghaleb2022studying}.
\emph{Finished earlier builds} counts the builds of the same CI workflow and branch that finished before the target build was created. We include this feature because the other history features are less reliable when they are computed from only a few earlier builds. Prior work on CI build prediction calls this situation the cold-start problem~\citep{jin2023hybridcisave}. With this count, a model can tell whether the other history features come from many earlier builds or from only a few, for example in the first builds of a new branch.
\emph{Hours since last finished build} is the time from the end of the most recently finished earlier build to the creation of the target build. We compute this feature for the same project and for the same CI workflow and branch. GitHub Actions can cache dependencies, such as installed Python packages, and reuse them in later builds. We include this feature because GitHub Actions removes caches that have not been accessed for more than seven days~\citep{github_actions_caching}. A build that starts after a long pause may therefore need to download its dependencies again. Prior work shows that downloading the same dependencies again at the start of each build wastes build time~\citep{gallaba2020accelerating}.
\emph{Running builds of the same CI workflow} counts the builds of the same CI workflow that were created before the target build and had not finished when the target build was created. We include this feature because these builds compete with the target build for the same runners. Prior work suggests that the workload of the CI service when a build is triggered can delay the build~\citep{laukkanen2015build,ghaleb2019empirical}.
When no build of a group has finished before the target build, the history features of that group have no value, and we set them to 0.

\begin{table}[htbp]
\centering
\caption[Feature families for CI build time prediction]{High-level overview of feature families and their member features for CI build time prediction.}
\label{tab:feature_families}
\scriptsize
\setlength{\tabcolsep}{5pt}
\renewcommand{\arraystretch}{1.1}
\begin{tabularx}{\linewidth}{@{}>{\raggedright\arraybackslash}p{2.6cm}>{\raggedright\arraybackslash}X>{\raggedright\arraybackslash}X@{}}
\toprule
\textbf{Feature family} & \textbf{Features} & \textbf{Rationale} \\
\midrule

\rowcolor{gray!10}
\multicolumn{3}{l}{\textbf{Run-level features}}\\[2pt]

(I) Code churn factors &
(1) Number of commits, (2) Added code lines, (3) Deleted code lines, (4) Total lines changed, (5) Modified source files, (6) Modified test files, (7) Modified build files &
Larger and more widely scattered pull request changes increase compilation and testing workload~\citep{ghaleb2019empirical,jin2020cost}. \\[2pt]

(II) Build history factors &
(1) Last build duration, same CI workflow and branch, (2) Last build duration, same CI workflow, (3) Median duration of last 20 builds, same project, (4) Duration variability of last 5 builds, same CI workflow and branch, (5) Failure rate of last 10 builds, same CI workflow and event, (6) Finished earlier builds, same CI workflow and branch, (7) Hours since last finished build, same project, (8) Hours since last finished build, same CI workflow and branch, (9) Running builds of the same CI workflow, (10) Builds since last failure &
The duration and outcome of earlier builds show how the same CI workflow has run before the target build~\citep{rausch2017empirical,chen2020buildfast}. \\[2pt]

(III) Build timing and context factors &
(1) Weekday, (2) Weekend, (3) Triggered by push, (4) On default branch &
Builds triggered at different times or by different events may face different CI load and run different jobs~\citep{ghaleb2019empirical}. \\[4pt]

\rowcolor{gray!10}
\multicolumn{3}{l}{\textbf{Workflow-level features}}\\[2pt]

(IV) DAG complexity factors &
(1) DAG node count, (2) DAG control flow complexity, (3) DAG coupling complexity, (4) Maximum steps per job &
These features capture the size of a CI workflow and the dependencies between its jobs~\citep{kaimann1974coefficient,mccabe1976complexity,cardoso2006discourse,vassallo2019automated,khatami2024catching}. \\[2pt]

(V) CI configuration factors &
(1) Has docker build step, (2) Uses cache, (3) Matrix operating system count, (4) Maximum matrix size &
Docker builds add setup work, caching can shorten repeated builds, and build matrices increase the number of job instances~\citep{openja2022studying,gallaba2020accelerating,bouzenia2024resource}. \\[4pt]

\rowcolor{gray!10}
\multicolumn{3}{l}{\textbf{Project-level features}}\\[2pt]

(VI) ML domain factors &
(1) Computer vision, (2) General purpose ML, (3) Scientific ML, (4) NLP, (5) Time series, (6) Speech / audio, (7) AutoML, (8) Graph ML, (9) Reinforcement learning, (10) Other &
Different ML domains rely on different hardware and data characteristics, leading to distinct CI patterns and resource usage profiles. \\[2pt]

(VII) ML usage depth and CI history factors &
(1) Framework / Infrastructure, (2) Research Application, (3) Library User, (4) Platform Tooling, (5) API Consumer, (6) Other, (7) Repo diversity, (8) Repo stability &
ML usage depth reflects how a repository implements or uses ML functionality, while prior CI history captures workflow breadth and historical stability~\citep{vasilescu2015quality}. \\

\bottomrule
\end{tabularx}
\end{table}

\paragraph*{(III) Build timing and context factors}
This family captures when a build is triggered during the week, which event triggers it, and whether it runs on the default branch.
We include the two timing features because prior work finds that builds triggered on weekdays are more likely to take long~\citep{ghaleb2019empirical}.
\emph{Weekday} is derived from the build start timestamp and encoded as an integer between 0 (Monday) and 6 (Sunday).
\emph{Weekend} is a binary feature that is set to 1 if the build starts on Saturday or Sunday, and 0 otherwise.
\emph{Triggered by push} is set to 1 if a push event triggers the build, and 0 otherwise. We include this feature because a CI workflow can run different jobs for different trigger events. For example, a CI workflow may run its packaging and release jobs only after a push~\citep{github_actions_workflow_syntax}.
\emph{On default branch} is set to 1 if the build runs on a branch named \texttt{main} or \texttt{master}, and 0 otherwise. We use these two names because they are the common names of the default branch on GitHub. We include this feature because a CI workflow can use branch filters to run some jobs only on specific branches~\citep{github_actions_workflow_syntax}. For example, a CI workflow may run its full test suite only on the default branch. Builds on the default branch may then run more jobs and take longer than builds on other branches.

\subsubsection{Workflow-Level}
\paragraph*{(IV) DAG complexity factors}
This family captures the structural complexity of the CI workflow by analyzing its job dependency graph.
We include these features because CI workflow structure can affect build duration by determining which jobs must wait and which jobs can run in parallel~\citep{lebeuf2018understanding,github_actions_jobs,github_actions_workflow_syntax}.
Prior work on workflow and process complexity also shows that graph structure can be used to measure branching and coupling~\citep{mccabe1976complexity,cardoso2006discourse}.
For each CI workflow YAML file, we construct a DAG where nodes represent jobs and edges represent \texttt{needs} dependencies.
\emph{DAG node count} records the number of jobs defined in the CI workflow YAML file.
The remaining DAG features summarize mechanisms that can affect CI build duration. At the CI workflow level, build duration is the elapsed time from the start of a CI workflow run until its final required job finishes. This time is not determined only by the number of jobs. It also depends on which jobs must wait for other jobs and where parallel branches of the workflow join. We therefore adapt established graph and process measures to CI workflows to capture these execution constraints~\citep{kaimann1974coefficient,mccabe1976complexity,cardoso2006discourse}. We do not assume that each feature has a fixed direction of effect. For example, more branches in a CI workflow let more jobs run at the same time and can shorten a build, but more joins make more jobs wait for other jobs and can lengthen a build. We use these features to test their association with the longest builds in RQ2 and their predictive value in RQ3.

\emph{DAG control flow complexity} measures the amount of branching and joining in the workflow graph. This matters for build duration because branches can create parallel work, while joins make a downstream job wait until all required upstream branches have finished. It is computed as $|\mathcal{E}| - |\mathcal{V}| + 2C$, where $|\mathcal{E}|$ and $|\mathcal{V}|$ are the numbers of edges and nodes, and $C$ is the number of weakly connected components. A weakly connected component is a maximal subgraph that remains connected when edge direction is ignored~\citep{mccabe1976complexity,lebeuf2018understanding}. In Figure~\ref{fig:dag-example}, the workflow has 7 jobs and 9 dependency edges, and it is a single component (\(C=1\)).
Thus the control flow complexity is \(9 - 7 + 2\times 1 = 4\), which reflects the branching from \icode{setup} into several parallel jobs that later join.

\emph{DAG coupling complexity} measures whether some jobs act as dependency hubs. This matters for build duration because a job with many incoming dependencies must wait for several upstream jobs, while a job with many outgoing dependents can delay several downstream jobs if it starts late or runs for a long time. Following Cardoso's definition of coupling~\citep{cardoso2006discourse}, we sum, for each job, the product of its number of incoming dependencies and its number of outgoing dependents.
In the example DAG in Figure~\ref{fig:dag-example}, the job \icode{model-evaluation} depends on three other jobs and is itself required by one job, so it contributes $3 \times 1 = 3$ to the coupling complexity.
The job \icode{code-quality} depends on one job and triggers one downstream job, so it contributes $1 \times 1 = 1$.
The \icode{data-validation}, \icode{model-training}, and \icode{model-testing} jobs each have one upstream and one downstream job, so together they contribute $1 + 1 + 1 = 3$ to the coupling complexity.
The remaining jobs either have no predecessors or no successors and therefore contribute $0$.
The overall \emph{DAG coupling complexity} of this example workflow is thus $3 + 1 + 3 = 7$.

\emph{Maximum steps per job} counts the largest number of steps within a single build job, based on the \texttt{steps} array in the YAML~\citep{khatami2024catching}. This feature is included because a build job with many steps can become a long running stage within the workflow.

\paragraph*{(V) CI configuration factors}
This family captures how a CI workflow builds Docker images, uses caching, and runs jobs under matrix configurations.
A matrix configuration runs the same job under several settings, such as different operating systems or Python versions.
We include \emph{Has docker build step} because building a Docker image can add setup work to a build. Docker images of ML projects often build on CUDA or ML framework images and contain many files, so they need more resources than the images of other software~\citep{openja2022studying}.
We parse each workflow YAML file, normalize it to lower case, and search for characteristic substrings.
\emph{Has docker build step} is set to 1 if the text contains ``\texttt{docker build}'', and 0 otherwise.
We include \emph{Uses cache} because caching can shorten a build by reusing dependencies from earlier builds~\citep{gallaba2020accelerating}.
We include \emph{Matrix operating system count} and \emph{Maximum matrix size} because a matrix configuration creates several job instances from one job definition and therefore adds work to a build.
However, GitHub Actions runs each job instance on its own runner, so matrix jobs can run in parallel and may not lengthen a build when enough runners are available~\citep{bouzenia2024resource}.
\emph{Uses cache} is set to 1 if the text contains ``\texttt{actions/cache}'', and 0 otherwise.
\emph{Matrix operating system count} counts the distinct values of the matrix keys whose names contain ``\texttt{os}'' over all jobs of the CI workflow, for example \texttt{os: [ubuntu-latest, windows-latest]}. When the CI workflow has no such key, we set this feature to 1, because its jobs then run on a single operating system.
\emph{Maximum matrix size} measures the largest matrix expansion within a workflow~\citep{wessel2023github}.
We first identify all jobs in the workflow that define a \texttt{strategy.matrix}.
For each such job, we compute its matrix size by multiplying the number of options in each matrix dimension.
For example, a job with a matrix over three operating systems and two Python versions creates $3 \times 2 = 6$ job instances, so the matrix size of that job is 6.
\emph{Maximum matrix size} is then the maximum of these per-job matrix sizes among all matrix-configured jobs in the same CI workflow.
\subsubsection{Project-Level}
\paragraph*{(VI) ML domain factors}
This family encodes the primary ML domain of each project in our studied set of ML repositories.
We include these features because different ML domains often require different data, model types, hardware, and validation tasks.
These differences can lead to different CI workflows and resource usage patterns.
For example, computer vision projects often depend on image datasets and GPU libraries, whereas NLP projects may load LLMs or language resources.
These domain labels come from the manual project classification conducted for RQ1.
In that process, three coders independently inspect each project's documentation and artifacts, including the project description, repository topics, and \texttt{README.md} file.
The coders then resolve disagreements through discussion.
We reuse the final consensus labels as project-level features in this prediction task, with the full coding protocol described in the approach for RQ1~in Section~\ref{approach:longbuilds:RQ1}.
The complete manual coding scheme contains ten ML-domain categories, which are reported in Table~\ref{tab:ml-domain-def}.
For the prediction feature set, we apply binary indicator encoding to convert the categorical ML-domain label into model inputs.
Specifically, we create one binary indicator for each of the ten ML-domain categories:
\emph{Computer vision}, \emph{General purpose ML}, \emph{Scientific ML}, \emph{NLP}, \emph{Time series}, \emph{ Speech/audio}, \emph{AutoML}, \emph{Graph ML}, \emph{Reinforcement learning}, and \emph{Other}.
For each repository, the feature corresponding to its primary ML domain is set to 1, and the remaining ML-domain features are set to 0.

\begingroup
\setlength{\extrarowheight}{0pt}
  \begin{table}[htbp]
    \footnotesize
    \centering
    \caption[ML domain categories for the 107 studied projects]{Categories used to code the primary ML domain for the 107 studied projects.}
    \label{tab:ml-domain-def}
    \begin{tabularx}{\linewidth}{@{}l>{\centering\arraybackslash}m{2.1cm}>{\raggedright\arraybackslash}X@{}}
      \toprule
      \textbf{Domain} & \makecell[c]{\textbf{Projects}} & \textbf{Description} \\
      \midrule
      General Purpose & 45 (42.1\%) &
      Reusable ML frameworks, toolkits, or libraries intended for several downstream tasks rather than a single application area. \\
      Scientific ML & 23 (21.5\%) &
      ML for scientific or medical problems, including molecular simulation (\eg~SchNetPack, TorchANI), drug discovery (\eg~DeepChem), materials informatics (\eg~SMACT), medical imaging (\eg~DIPY), and astronomy (\eg~astroNN). \\
      NLP & 10 (9.3\%) &
      Text and language applications, including generation, translation, summarization, and question answering. \\
      Computer Vision & 7 (6.5\%) &
      Image and video applications, including detection, segmentation, recognition, and other visual perception tasks. \\
      Time Series & 2 (1.9\%) &
      Temporal or sequential data applications, including forecasting, trading (\eg~OctoBot), and portfolio optimization (\eg~poptimizer). \\
      Speech / Audio & 2 (1.9\%) &
      Speech and audio applications, including recognition, enhancement, and music analysis. \\
      AutoML & 2 (1.9\%) &
      Systems for automating model selection, hyperparameter optimization, neural architecture search, or pipeline configuration. \\
      Graph ML & 2 (1.9\%) &
      Learning methods for graph data, including graph neural networks (\eg~DGL) and knowledge graph embedding (\eg~PyKEEN). \\
      Reinforcement Learning & 1 (0.9\%) &
      Learning problems where agents interact with an environment and optimize policies through rewards. \\
      Other & 4 (3.7\%) &
      Projects whose main application area is not clear enough to assign to the above domains. \\
      \bottomrule
    \end{tabularx}
  \end{table}
\endgroup

\paragraph*{(VII) ML usage depth and CI history factors}
This family captures project-level context that may shape CI workload before a build starts.
It combines two types of information.
First, ML usage depth describes how a repository implements or uses ML functionality, because different forms of ML use tend to require different CI tasks.
For example, framework and infrastructure projects may exercise low level ML components across several environments, whereas library user or platform tooling projects may run narrower checks around downstream use, orchestration, or deployment support.
Second, two CI-history features summarize the repository's prior workflow use and prior build outcomes before the target build~\citep{vasilescu2015quality}.
The ML usage depth labels are produced using the same manual classification process as the ML domain labels.
That is, the three coders independently inspect the project documents, compare their labels, and resolve disagreements through discussion before assigning a final consensus ML usage depth label to each repository.
Inter-rater agreement for the ML usage depth labels is high, with Fleiss' kappa of 0.924.
Raw percentage agreement can overstate reliability because coders may agree on some labels by chance, particularly when a few categories are much more frequent than others.
Cohen's kappa corrects for chance agreement, but it is defined for two raters only.
Fleiss' kappa extends this correction to settings in which more than two raters assign each item to one of several nominal categories~\citep{fleiss1971measuring}.
This matches our coding setting, in which three coders independently assign each repository to one of six ML usage depth categories.
Fleiss' kappa has also been used in prior work on CI in ML projects, where three authors independently categorized the tasks executed in CI configuration files~\citep{rzig2022characterizing}.
We interpret the values following \citet{landis1977measurement}, where values above 0.80 indicate almost perfect agreement.
We reuse these final labels as project-level features in this prediction task, with the full coding protocol described in the approach for RQ1~in Section~\ref{approach:longbuilds:RQ1}.
The complete manual coding scheme contains six ML usage depth categories, which are reported in Table~\ref{tab:usage-depth-def}.
For the prediction feature set, we apply binary indicator encoding to convert the categorical ML usage depth label into model inputs.
Specifically, we create one binary indicator for each of the six ML usage depth categories:
\emph{ Framework/Infrastructure}, \emph{Research Application}, \emph{Library User}, \emph{Platform Tooling}, \emph{API Consumer}, and \emph{Other}.
For each repository, the feature corresponding to its ML usage depth category is set to 1, and the remaining ML usage depth features are set to 0.

\begingroup
\setlength{\extrarowheight}{0pt}
\begin{table}[htbp]
    \small
    \centering
    \caption{Categories used to code ML usage depth for the 107 studied projects.}
    \label{tab:usage-depth-def}
    \begin{tabularx}{\linewidth}{@{}l>{\centering\arraybackslash}m{2.1cm}>{\raggedright\arraybackslash}X@{}}
      \toprule
      \textbf{Category} & \makecell[c]{\textbf{Projects}} & \textbf{Description} \\
      \midrule
      Framework / Infrastructure & 77 (72.0\%) &
      Core ML frameworks or infrastructure, including tensor operations, automatic differentiation, optimizers, neural network layers, and related runtime components. \\
      Platform Tooling & 9 (8.4\%) &
      Tools for ML pipelines and operations, such as orchestration, model serving, experiment management, and deployment. Training or inference is not the main workload. \\
      Library User & 8 (7.5\%) &
      Applications that use existing ML libraries for training or inference rather than implementing the learning method itself. Many adapt pretrained models or library APIs to a downstream task.\\
      Research Application & 4 (3.7\%) &
      Research codebases for a specific ML problem, usually including model code, training scripts, evaluation code, and sometimes custom losses or architectures. \\
      API Consumer & 1 (0.9\%) &
      Applications that call external ML services or hosted model APIs without local model training or inference. \\
      Other & 8 (7.5\%) &
      Projects whose main use of ML is not clear enough to assign to the above categories. \\
      \bottomrule
    \end{tabularx}
  \end{table}
\endgroup

\emph{Repo diversity} is computed per build as the number of distinct workflows that the repository has used before the current build, based on unique workflow identifiers in its past runs.
We include this feature because a repository that has used more workflows has exposed more CI entry points, which may correspond to broader CI coverage across tasks such as testing, packaging, documentation, or deployment.
Such breadth can be related to build duration because repositories with broader CI usage may contain workflows with more varied workloads and resource requirements.

\emph{Repo stability} is the historical success rate of a repository up to the current build.
For each build, we sort all runs of that repository by timestamp and compute the success rate using only the runs that occurred before the current one.
We include this feature because lower historical success rates may indicate recurring CI problems, such as flaky tests, dependency issues, or unstable workflow configurations, that can persist across builds and affect build duration or reliability~\citep{rausch2017empirical,ghaleb2019studying,chen2020buildfast}.

\section{Research Questions and Findings}
\label{sec:longbuilds:rqs}
\subsection{RQ1: How can the longest builds in ML projects be characterized?}
\label{approach:longbuilds:RQ1}

\paragraph{Motivation}
A CI build is designed to provide rapid feedback to developers. However, in the context of ML projects, CI workflows may involve computationally expensive tasks such as model training, large-scale data processing, and integration testing across diverse hardware environments.
These CI workflow characteristics raise the possibility that build time may be unevenly distributed across builds, with some builds creating much longer feedback delays than others.
This motivates us to examine whether a relatively small fraction of builds is responsible for most of the total time spent on CI builds in ML projects.
For this paper, we therefore focus on the longest builds that consume much of the total time spent on CI builds.
We make this choice because our goal is to identify the builds that take up most of the overall build time and are therefore most relevant for proactive CI optimization.
This differs from prior threshold based modeling of long build durations, where long builds are defined relative to project specific quantiles~\citep{ghaleb2019empirical}.
Furthermore, we seek to understand whether these longest builds are uniformly distributed or concentrated within specific types of ML projects or specific domains. This analysis helps identify which segments of the ML ecosystem face the most severe CI bottlenecks.
In addition, we investigate whether the longest builds exhibit a significant difference across ML projects.
Specifically, we examine their distribution along two independent project dimensions. The first is \textit{ML domain}, which describes the main application area of a project (\eg~computer vision, NLP, or scientific ML). The second is \textit{ML usage depth}, which describes how a project engages with ML, for example by calling external ML services, training or fine-tuning models, or providing ML infrastructure. Each project receives one label on each dimension, so the two dimensions describe different aspects of the same project rather than a single combined category system. We use these two dimensions to identify where CI bottlenecks in ML projects are most significant.

\paragraph{Approach}
To address RQ1, we analyze all collected builds using a four-step methodology.
First, we examine how build time is concentrated across all builds. We sort all builds by duration in descending order, compute the cumulative sum of durations, and calculate the cumulative percentage of the total time spent on CI builds as the number of builds increases.
We then identify the smallest top percentile of builds whose cumulative duration reaches 80\% of that total time.
In our dataset, this corresponds to the top 18\% longest builds, \ie~builds that take at least 39.20 minutes. We use these longest builds for the remainder of the paper because this threshold is derived from the observed concentration of CI time. This choice therefore aligns the analysis with our practical goal of identifying the builds that create the largest CI time burden. In contrast, prior work such as \citet{ghaleb2019empirical} uses threshold scenarios, including an above median threshold, to compare short and long builds within each project. That design is suitable for studying general duration differences, but it is less suitable for our goal because it would label about half of all builds as long. Such a broad target group would include many builds that contribute relatively little to the total CI time and would dilute the upper tail builds that are most relevant for CI optimization. Our top 18\% definition is therefore intended to isolate the builds that dominate total CI time, rather than to divide the full build population into broad short and long groups.

Second, we examine project heterogeneity in build time concentration.
We do this step because the global top 18\% threshold is derived from all builds, whereas the meaning of a longest build may differ in each project.
We calculate two project-level indicators: (1) the percentage of total build time contributed by the top 18\% longest builds in that project, and (2) the percentage of builds needed to reach 80\% of that project's total build time.
These indicators allow us to assess whether the concentration pattern observed across all studied projects is also present within individual projects, and how strongly it varies from one project to another.

Third, we examine how these longest builds are distributed across different types of ML projects by manually classifying all 107 projects.
We follow a document analysis approach~\citep{bowen2009document}, which codes project documents into themes.
The input materials (\ie~project documents) include project descriptions, repository topics, and README files.
We classify each project along two dimensions, \textit{ML usage depth} and \textit{ML domain}.
We manually label each project on the two dimensions separately.
\textit{ML usage depth} captures the extent to which a project engages with ML, ranging from projects that mainly consume external model APIs to projects that train~and fine-tune models, or provide ML infrastructure.
\textit{ML domain} denotes the primary application area, such as computer vision, speech/audio, or NLP.
To analyze the categories of the 107 ML projects, we employ an
inductive thematic analysis, which is designed for identifying, analyzing, and reporting themes within qualitative data, following the guidelines by \citet{nowell2017thematic}.
The document analysis uses the following procedure.
We first collect the project documents for the 107 projects via the GitHub API, including project descriptions, repository topics, and \texttt{README.md} files.
Three coders then independently code the same document set along two separate dimensions, \ie~ML usage depth and ML domain. One coder is the first author, and the other two coders are doctoral researchers. Inter-rater agreement on the initial independent labels is high, with Fleiss' kappa of 0.964 for the ML-domain labels and 0.924 for the ML usage depth labels.
After this independent coding round, the three coders compare their labels, discuss disagreements, and refine the coding scheme where necessary until they reach final consensus labels for all projects.
It is important to note that this document analysis aims to broaden our understanding of the categories represented in the 107 ML projects we study and how different types of ML projects experience the longest builds.
The classified project categories and their associated build-duration distributions are reported in the findings for RQ1 below.

Finally, we test whether build durations differ across ML project categories.
For this analysis, we compare the observed build durations across the two classification dimensions, \ie~\textit{ML usage depth} and \textit{ML domain}.
We apply a Kruskal--Wallis H test to assess whether build-duration distributions significantly differ across these project categories.
The null hypothesis of this test is that the build-duration ranks are similarly distributed across project categories.
If the Kruskal--Wallis test is significant with $p$-value $< 0.05$, we reject the null~hypothesis and infer that at least one category differs from the others.
We then perform pairwise Mann--Whitney U tests to localize the differences.
For each category pair, the null hypothesis is that projects in the two categories of ML usage depth (or ML domains) share the same build-duration distribution.
A $p$-value below 0.05 leads us to reject this null hypothesis, indicating a statistically significant difference in build duration between the two tested categories of ML projects.
These comparisons identify which two specific ML project categories exhibit statistically significant differences in build duration.
After the Mann--Whitney U tests, we also report Cliff's delta as a nonparametric effect size to quantify differences between two build duration distributions.
Cliff's delta reflects the extent to which durations from one category tend to be larger than those from another.
We interpret the absolute value using standard thresholds.
Values below 0.147 are negligible, 0.147 to 0.33 are small, 0.33 to 0.474 are medium, and greater than 0.474 are large~\citep{romano2006appropriate,torchiano2016effsize}.
A positive delta means the first category typically has longer build duration, while a negative delta indicates the opposite.

\paragraph{Findings}

\begin{table}[htbp]
  \small
  \centering
  \caption[Build-time concentration across thresholds]{Build-time concentration and failure characteristics across different thresholds (total builds $N = 269{,}930$ in 107 ML projects).}
  \label{tab:long-builds-overview}
  \begin{tabularx}{\linewidth}{%
      l
      >{\raggedleft\arraybackslash}X
      >{\centering\arraybackslash}X
      >{\centering\arraybackslash}X
      >{\raggedleft\arraybackslash}X
      >{\raggedleft\arraybackslash}X
      >{\raggedleft\arraybackslash}X}
    \toprule
    & & \multicolumn{2}{c}{\textbf{Duration (min)}} & & & \\
    \cmidrule(lr){3-4}
    \textbf{Threshold}
      & \textbf{Count}
      & \textbf{Mean}
      & \textbf{Median}
      & \textbf{Failure rate}
      & \textbf{Time} \%
      & \textbf{Pareto ratio} \\
    \midrule
    Top 5\% & 13{,}497 & 416.79 & 271.07 & 39.99\% & 53.80\% & 10.76$\times$ \\
    Top 10\% & 26{,}994 & 265.69 & 169.16 & 43.03\% & 68.59\% & 6.86$\times$ \\
    Top 15\% & 40{,}498 & 198.42 & 111.31 & 40.15\% & 76.85\% & 5.12$\times$ \\
    \textbf{Top 18\%} & \textbf{48{,}605} & \textbf{172.77} & \textbf{87.73} & \textbf{39.09\%} & \textbf{80.30\%} & \textbf{4.46$\times$} \\
    Top 20\% & 53{,}993 & 159.15 & 78.77 & 39.14\% & 82.17\% & 4.11$\times$ \\
    \bottomrule
  \end{tabularx}
\end{table}

\vspace{0.5em}
\noindent\textbf{Finding 1: The top 18\% longest builds account for about 80\% of total build time.}
As shown in Table~\ref{tab:long-builds-overview}, across all 269{,}930 builds from our 107 studied projects, the top 18\% longest builds (48{,}605 builds, each taking at least 39.20 minutes) account for 80.30\% of the total build time, with a median duration of 87.73 minutes and a failure rate of 39.09\%.
This top 18\% slice includes builds from 78 of the 107 projects in our dataset, so the concentration is not driven only by a very small number of repositories.
This finding is consistent with the Pareto principle (the 80/20 rule)~\citep{fenton2014software} when all builds are considered together.
Tightening the threshold to the top 10\% and top 5\% of builds still yields 68.59\% and 53.80\% of total time, respectively, while relaxing it to 20\% increases the time share only modestly to 82.17\%.

\begin{figure}[tbp]
  \centering
  \includegraphics[width=\linewidth]{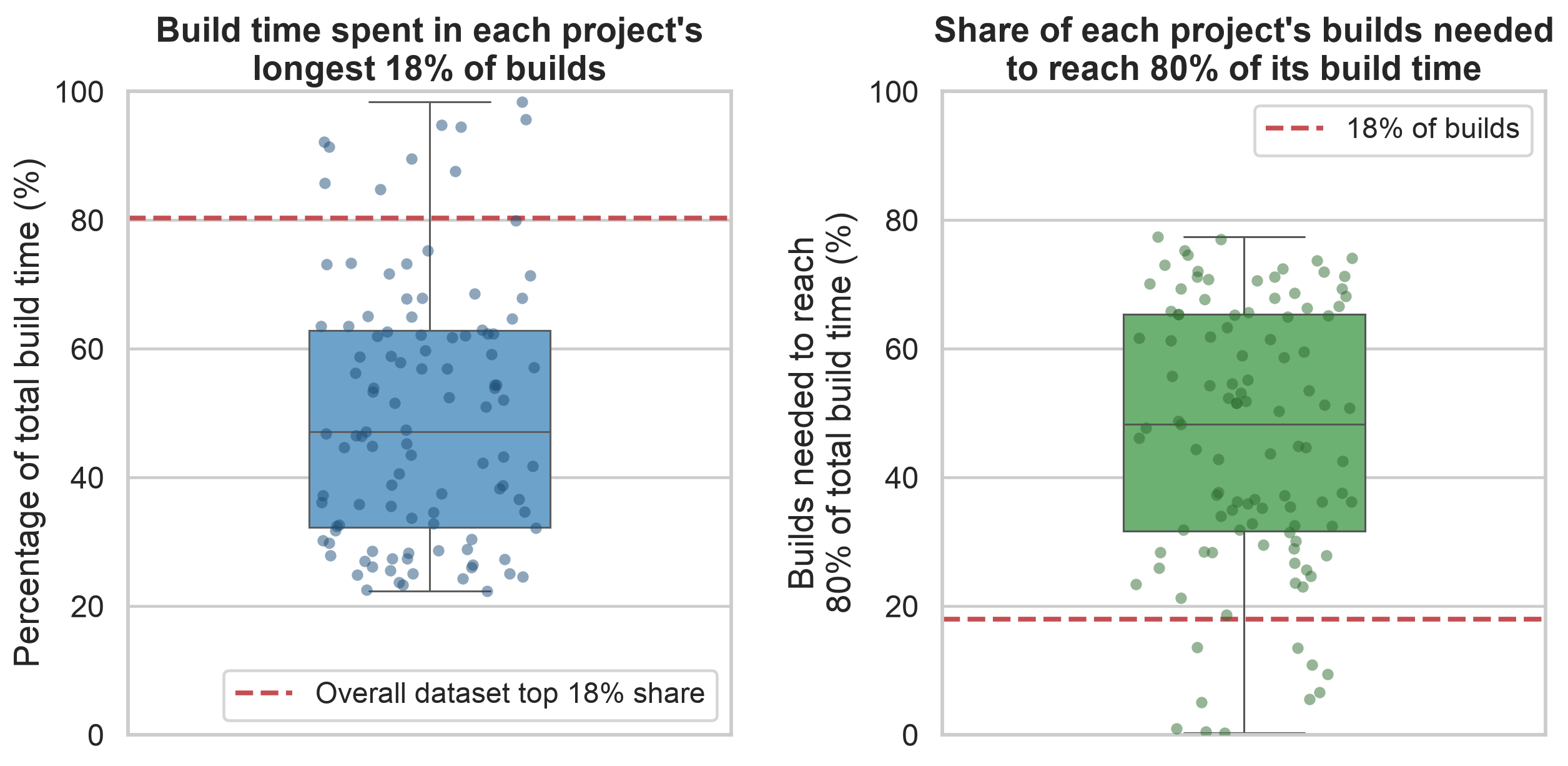}
  \caption{Project-level heterogeneity in build-time concentration.}
  \caption*{\footnotesize Note: Each point represents one of the 107 studied projects. The left boxplot shows, for each project, the percentage of total build time contributed by its top 18\% longest builds. The right boxplot shows, for each project, the percentage of builds needed to reach 80\% of that project's total build time.}
  \label{fig:project_level_concentration}
\end{figure}

\vspace{0.5em}
\noindent\textbf{Finding 2: In about half of the projects, the long builds of the project take more than half of its total build time.}
Figure~\ref{fig:project_level_concentration} shows how much of the build time of each project goes to the longest 18\% of builds of that project.
In at least half of the studied projects (54 of 107), the top 18\% longest builds account 
for 71.8\% or more of the project's total build time.
In 75\% of the projects, the top 18\% longest builds account for at least 55.0\% of the 
project's total build time.
The right boxplot in Figure~\ref{fig:project_level_concentration} shows the same pattern 
from another angle. In at least half of the projects, only the top 18.1\% of longest builds 
are needed to reach 80\% of the project's total build time.
Together, these results show that the longest builds identified by the Pareto-based threshold (\ie~the top 18\% longest builds) also contribute a large build time within individual projects, although the degree of concentration differs across projects.

\begin{figure}[htbp]
  \centering
  \begin{subfigure}{\textwidth}
    \centering
    \includegraphics[width=\linewidth]{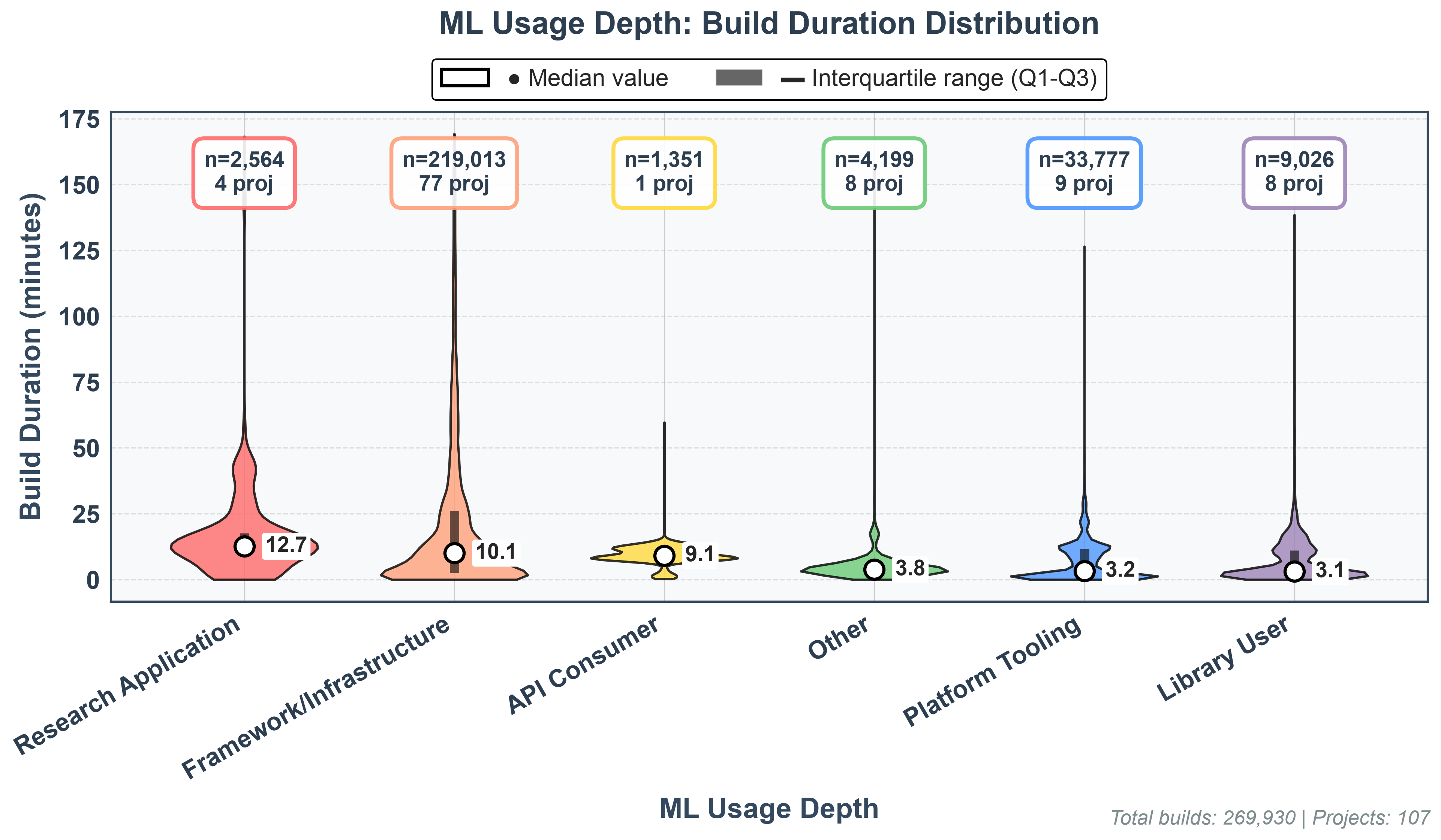}
    \caption{ML Usage Depth}
    \label{fig:ml_usage_depth}
  \end{subfigure}
  \hfill
  \begin{subfigure}{\textwidth}
    \centering
    \includegraphics[width=\linewidth]{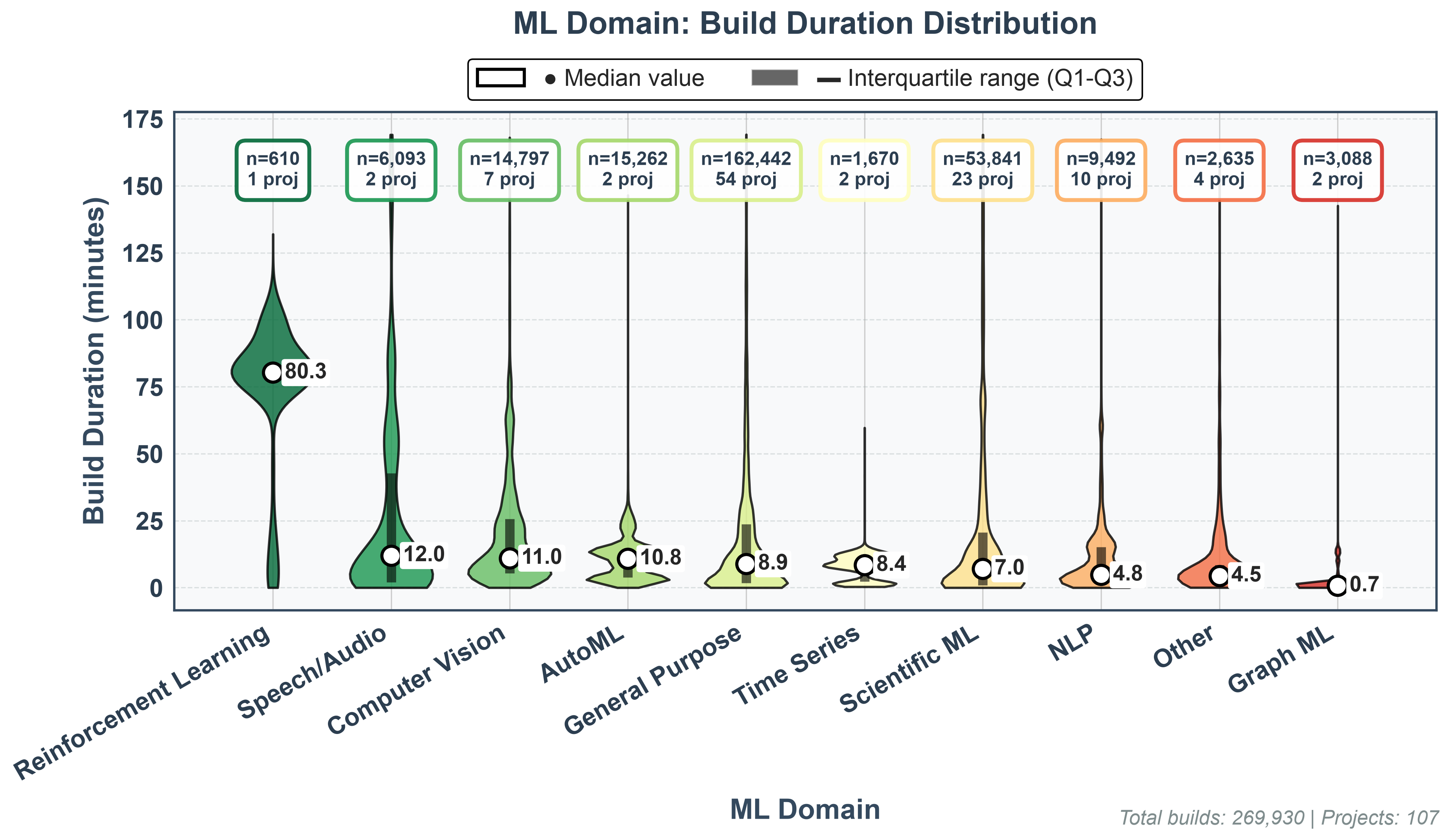}
    \caption{ML Domain Distribution}
    \label{fig:ml_domain}
  \end{subfigure}
  \caption{Overview of machine learning usage depth and domain distribution.}
  \label{fig:ml_combined}
\end{figure}

\vspace{0.5em}
\noindent\textbf{Finding 3: Build duration differs across ML usage depth categories, and ML framework and infrastructure projects produce almost all of the top 18\% longest builds.}
Table~\ref{tab:usage-depth} and Figure~\ref{fig:ml_usage_depth} show that \emph{Framework/Infrastructure} projects, which implement low-level ML libraries such as tensor operations, autograd, and optimizers, contribute 219{,}013 builds from 77 projects (81.1\% of all builds), with a median duration of 10.98 minutes.
Their builds also have the longest upper tail, with a 95th percentile of 207.88 minutes, the highest of all categories.
As a result, 47{,}808 of the 48{,}605 top 18\% longest builds (98.4\%) come from \emph{Framework/Infrastructure} projects, and 21.8\% of their builds fall in the top 18\%, compared with at most 10.5\% in any other category.
In contrast, \emph{Library User} and \emph{Platform Tooling} projects mainly rely on existing ML libraries or focus on supporting ML pipelines and deployment workflows,~and have shorter builds with medians of 3.08 and 3.23 minutes, respectively.
A Kruskal--Wallis test over 269{,}930 builds confirms significant differences in build duration across ML usage depth categories ($p < 0.001$).
Pairwise Mann--Whitney U tests show that \emph{Framework/Infrastructure} builds are significantly longer than \emph{Library User} builds and builds of projects labeled \emph{Other}, with small effect sizes (Cliff's $\delta = 0.32$ for both), and longer than \emph{Platform Tooling} builds, with a medium effect size ($\delta = 0.35$).
\emph{Research Application} projects have the highest median (12.93 minutes), and the single \emph{API Consumer} project has a median of 9.13 minutes.
However, these two categories contain only four projects and one project, respectively, and their differences from \emph{Framework/Infrastructure} are negligible ($\delta = -0.05$ and $0.11$).
Overall, these results indicate that ML framework and infrastructure projects have longer builds than projects that consume ML through libraries or MLOps tooling, and that they account for almost all of the longest builds.

\begin{table}[htbp]
  \small
  \centering
  \caption{Build duration by ML usage depth (269{,}930 builds with depth labels).}
  \label{tab:usage-depth}
  \begin{tabularx}{\linewidth}{l
      >{\raggedleft\arraybackslash}X
      >{\raggedleft\arraybackslash}X
      >{\raggedleft\arraybackslash}X
      >{\raggedleft\arraybackslash}X}
    \toprule
    \textbf{Category} & \textbf{Builds} & \textbf{Projects} & \textbf{Median (min)} & \textbf{P95 (min)}$^{*}$ \\
    \midrule
    Research Application & 2{,}564 & 4 & 12.93 & 44.74 \\
    Framework / Infrastructure & 219{,}013 & 77 & 10.98 & 207.88 \\
    API Consumer & 1{,}351 & 1 & 9.13 & 13.67 \\
    Platform Tooling & 33{,}777 & 9 & 3.23 & 22.20 \\
    Library User & 9{,}026 & 8 & 3.08 & 21.62 \\
    Other & 4{,}199 & 8 & 3.98 & 59.12 \\
    \bottomrule
  \end{tabularx}
  \par\vspace{2pt}
  {\raggedright\scriptsize Note: $^{*}$ \textbf{P95} is the 95th percentile of build duration.\par}
\end{table}

\begin{table}[!htbp]
  \small
  \centering
  \caption{Summary statistics of build duration across ML domain categories.}
  \label{tab:ml-domain}
  \begin{tabularx}{\linewidth}{l
      >{\raggedleft\arraybackslash}X
      >{\raggedleft\arraybackslash}X
      >{\raggedleft\arraybackslash}X
      >{\raggedleft\arraybackslash}X}
    \toprule
    \textbf{Domain} & \textbf{Builds} & \textbf{Projects} & \textbf{Median (min)} & \textbf{P95 (min)}$^{*}$ \\
    \midrule
    Reinforcement Learning & 610 & 1 & 80.31 & 100.24 \\
    Speech / Audio & 6{,}093 & 2 & 12.13 & 146.37 \\
    Computer Vision & 14{,}797 & 7 & 11.32 & 74.91 \\
    AutoML & 15{,}262 & 2 & 10.83 & 25.32 \\
    General Purpose & 162{,}442 & 54 & 10.02 & 193.48 \\
    Scientific ML & 53{,}841 & 23 & 8.57 & 214.60 \\
    Time Series & 1{,}670 & 2 & 8.41 & 13.53 \\
    NLP & 9{,}492 & 10 & 5.29 & 111.94 \\
    Graph ML & 3{,}088 & 2 & 0.73 & 12.78 \\
    Other & 2{,}635 & 4 & 4.85 & 240.51 \\
    \bottomrule
  \end{tabularx}
  \par\vspace{2pt}
  {\raggedright\scriptsize Note: $^{*}$ \textbf{P95} is the 95th percentile of build duration.\par}
\end{table}

\vspace{0.5em}
\noindent\textbf{Finding 4: Build duration varies significantly across ML domains, but most differences between the larger domains are small or negligible.}
Table~\ref{tab:ml-domain} and Figure~\ref{fig:ml_domain} show variation across ML domains.
In our sample, the single \emph{Reinforcement Learning} (RL) project has the highest median build duration at 80.31 minutes.
Because this category contains only one project, we treat this result as a descriptive observation rather than evidence that RL projects generally have longer CI builds.
One possible explanation is that, when CI workflows include RL-specific tests or simulations, they may inherit some of the computational demands of RL workloads, such as repeated environment interaction and model evaluation~\citep{mnih2015human,espeholt2018impala}.
Among the four domains with at least seven projects, \emph{Computer Vision} has the highest median (11.32 minutes), followed by \emph{General Purpose} (10.02 minutes), \emph{Scientific ML} (8.57 minutes), and \emph{NLP} (5.29 minutes).
\emph{Computer Vision} builds are longer than \emph{NLP} builds with a small effect size (Cliff's $\delta = 0.29$), and \emph{Scientific ML} builds are shorter than \emph{Computer Vision} builds with a small effect size ($\delta = -0.15$), whereas the differences between \emph{Computer Vision} and \emph{General Purpose} ($\delta = 0.11$) and among \emph{General Purpose}, \emph{Scientific ML}, and \emph{NLP} ($|\delta| \leq 0.11$) are negligible.
The upper tails differ more than the medians.
\emph{Scientific ML} and \emph{General Purpose} builds have the highest 95th percentiles among these four domains (214.60 and 193.48 minutes), and together they contribute 87.7\% of the top 18\% longest builds (24.1\% and 63.6\%, respectively).
\emph{Scientific ML} also has the largest proportion of its own builds in the top 18\% (21.7\%), compared with 19.0\% for \emph{General Purpose}, 16.3\% for \emph{Computer Vision}, and 9.7\% for \emph{NLP}.
By contrast, \emph{Graph ML} builds,~from only two projects, have a median duration of less than one minute and are shorter than the builds of every other domain, with large effect sizes ($|\delta| \geq 0.56$).
The Kruskal--Wallis test across domains indicates significant differences in build duration ($p < 0.001$).
Overall, these findings show that the top 18\% longest builds are not uniformly distributed across ML projects. Instead, they are concentrated in \emph{Framework/Infrastructure} projects on the ML usage depth dimension and in \emph{General Purpose} and \emph{Scientific ML} projects on the ML domain dimension.

\begin{tcolorbox}[colback=gray!5!white, colframe=gray!75!black, title=Summary of RQ1, markedbreak]
  The top 18\% longest builds account for approximately 80\% of total build time across the 269{,}930 studied builds, showing a Pareto style concentration.
  This concentration pattern is also present within individual project, although its strength varies across projects.
  Build duration differs significantly across ML usage depth categories and ML domain categories.
  Framework/Infrastructure projects have longer builds than Library User and Platform Tooling projects and contribute 98.4\% of the top 18\% longest builds, while Library User projects have the lowest median build duration.
  Across ML domains, most differences between the larger domains are small or negligible. General Purpose and Scientific ML projects contribute most of the longest builds, whereas Graph ML builds are much shorter than those of all other domains.
  \end{tcolorbox}

\subsection{RQ2: How are CI workflow characteristics associated with the longest build durations in ML projects?}
\paragraph{Motivation}
RQ1 shows that, in our dataset, the longest 18\% of builds account for most of total build time and are not distributed uniformly across ML usage depth categories and ML domains.
However, these project level categories do not explain which properties of the CI workflows themselves are associated with long build duration.
We therefore move from project characteristics to CI workflow characteristics in RQ2.

In GitHub Actions, a CI workflow run consists of jobs connected by dependency constraints.
Some jobs can run in parallel, while others must wait for upstream jobs to finish.
As a result, build duration depends not only on the runtime of each job, but also on which jobs must wait for other jobs.
This motivates our analysis of CI workflow structure, including the number of jobs, the dependencies between jobs, and matrix configurations.
Prior work on build duration has mainly characterized build time using traditional software engineering features, such as code churn, commit size, and other code-level change measures.
However, ML projects differ from general software projects because their CI workflows often include data preparation, model training, evaluation, and specialized execution environments, all of which can shape build time through workflow design rather than through code-change characteristics alone.
More specifically, a single ML CI workflow may include data preparation, model training, evaluation, packaging, and testing, with explicit dependencies among these jobs.
ML projects may also require specialized execution environments, such as third-party library dependencies, GPU hardware, or multi-platform tests.
These requirements can make CI workflow definitions more complex and can lengthen build time.
Against this background, RQ2 examines whether differences in CI workflow structure and configuration are associated with the longest 18\% of builds in ML projects.

\paragraph{Approach}
We compare CI workflow characteristics between the longest 18\% of builds and the remaining builds.
First, we leverage the 289 CI workflow DAGs constructed from the unique GitHub CI workflow files associated with our 107 studied projects, as described in Section~\ref{subsec:longbuilds:data_collection}.
Each DAG captures job dependencies extracted from a CI workflow YAML file.
Using these CI workflow DAGs and their associated YAML files, we analyze CI workflow related features that are defined earlier in Section~\ref{subsec:longbuilds:feature_engineering}. These features describe how large the CI workflow is, how jobs depend on one another, how much of the CI workflow must run sequentially or can run in parallel, and how the CI workflow is configured to run.
We split the builds into two groups, \ie~the longest 18\% of builds and the remaining builds, and then compare the distribution of each CI workflow feature between these two groups of builds.
For each continuous CI workflow feature, we use the Mann--Whitney U test to examine whether its distribution differs between the longest 18\% of builds and the remaining builds. We also report Cohen's~ $d$ effect sizes to distinguish statistical significance from practical importance. Effect sizes are interpreted following standard conventions used in prior work~\citep{cohen2016power,martens2019towards,dell2023evaluating}: negligible ($|d| < 0.2$), small ($0.2 \leq |d| < 0.5$), medium ($0.5 \leq |d| < 0.8$), and large ($|d| \geq 0.8$).
For categorical CI workflow features, such as \emph{Has docker build step} and \emph{Uses cache}, we use chi-square tests of independence.
To understand what types of CI workflows are more likely to produce the longest builds, we manually classify the 289 CI workflows into seven CI workflow categories, as defined in Table~\ref{tab:rq2-workflow-categories}.
We assign these categories by inspecting \texttt{workflow} names, \texttt{job} names, \texttt{step} names, and CI workflow YAML configuration content.
Each CI workflow is assigned to one category based on its main purpose. For example, a CI workflow that builds and tests the project only for one target platform is labeled \texttt{Platform-Specific} rather than \texttt{Build \& Packaging} or \texttt{Testing}.
Three coders independently label the same set of CI workflows using these category definitions.
The initial pairwise agreement between each pair of coders ranges from 98.6\% to 99.3\%.
 As for the ML usage depth labels, we also report Fleiss' $\kappa$.
Fleiss' $\kappa$ is 0.987, indicating high agreement among the coders.
The coders then discuss disagreements and refine the category definitions where needed until a consensus label is assigned to each CI workflow.
We then compare how often the longest 18\% of builds occur across these CI workflow categories.
This analysis shows which CI workflow characteristics are associated with the longest 18\% of builds in ML projects.

\begin{table}[htbp]
  \small
  \centering
  \caption[CI workflow categories]{Definitions of the CI workflow categories used to manually classify the 289 CI workflows.}
  \label{tab:rq2-workflow-categories}
  \begin{tabularx}{\linewidth}{l
      >{\centering\arraybackslash}p{0.14\linewidth}
      >{\raggedright\arraybackslash}X}
    \toprule
    \textbf{Category} & \textbf{Count (\%)} & \textbf{Description} \\
    \midrule
    Build \& Packaging & 27 (9.3\%) & Compiles source code, creates artifacts (e.g., wheels, binaries, jar), and builds across platforms (e.g., \texttt{linux}, \texttt{windows}, \texttt{macOS}). \\ \midrule
    Testing & 137 (47.4\%) & Executes unit tests, integration tests, and benchmarks to validate code correctness. \\ \midrule
    Code Lint & 32 (11.1\%) & Performs static code analysis, style checking, and formatting validation. \\ \midrule
    Platform-Specific & 66 (22.8\%) & Builds and tests the project for one specific operating system, compiler, or hardware architecture (e.g., \texttt{linux}, \texttt{clang-sanitizers}, \texttt{gcc-build}, \texttt{riscv}), rather than across platforms. \\ \midrule
    GPU \& CUDA & 11 (3.8\%) & Runs CI workflows requiring GPU acceleration or CUDA computing. \\ \midrule
    Documentation & 6 (2.1\%) & Generates project documentation and manages release processes. \\ \midrule
    Framework-Specific & 10 (3.5\%) & Integrates with particular ML frameworks or tools (e.g., XGBoost, PyTorch, TensorFlow). \\
    \bottomrule
  \end{tabularx}
\end{table}

\paragraph{Findings}

\vspace{0.5em}
\noindent\textbf{Finding 1: The longest 18\% of builds are most strongly associated with the duration of earlier builds of the same CI workflow.} As Table~\ref{tab:rq2-dag-comparison} shows, the build history features are the only features with large effect sizes.
The previous build of the same CI workflow and branch took 138.60 minutes on average for the longest 18\% of builds, compared with 17.92 minutes for the remaining builds.
The same pattern holds at the project level. The median duration of the last 20 builds of the project is 73.92 minutes for the longest 18\% of builds and 15.92 minutes for the remaining builds.
The long builds (the top 18\% of builds) also start while more builds of the same CI workflow are still running (1.77 vs.\ 0.52).

\vspace{0.5em}
\noindent\textbf{Finding 2: The structure of a CI workflow is more consistently associated with the longest 18\% of builds than its configuration.} As Table~\ref{tab:rq2-dag-comparison} shows, all four DAG structure metrics differ significantly between the two groups ($p < 0.001$), and all four have medium effect sizes.
DAG coupling complexity shows the largest relative difference. Its mean is about seven times higher for the longest 18\% of builds than for the remaining builds (5.60 vs.\ 0.79).
In the CI workflows of these builds, many jobs wait for several earlier jobs and are needed by several later jobs.
A job cannot start until all the jobs it needs have finished. These waits may make the whole build longer.

\begin{table}[htbp]
  \footnotesize\linespread{1.0}\selectfont
  \setlength{\tabcolsep}{4pt}
  \centering
  \caption[Workflow and build history features of the longest 18\% of builds]{CI workflow and build history features for the longest 18\% of builds \textbf{vs.} the remaining builds.}
  \label{tab:rq2-dag-comparison}
  \begin{tabularx}{\linewidth}{%
      >{\raggedright\arraybackslash}X
      >{\centering\arraybackslash}p{1.2cm}
      >{\centering\arraybackslash}p{1.5cm}
      >{\centering\arraybackslash}p{1.1cm}
      >{\centering\arraybackslash}p{0.5cm}
      >{\centering\arraybackslash}p{1.4cm}}
    \toprule
    \textbf{Metric}
      & \textbf{Longest 18\%}
      & \textbf{Remaining}
      & \textbf{Diff. (\%)}
      & \textbf{$p$}
      & \textbf{Effect size} \\
    \midrule
    \multicolumn{6}{c}{\textbf{CI Workflow DAG Structure}} \\
    \midrule
    DAG node count
      & 9.49
      & 2.66
      & +256.5
      & ***
      & Medium \\
    DAG control flow complexity
      & 3.26
      & 1.79
      & +81.9
      & ***
      & Medium \\
    DAG coupling complexity
      & 5.60
      & 0.79
      & +612.2
      & ***
      & Medium \\
    Maximum steps per job
      & 10.05
      & 7.37
      & +36.4
      & ***
      & Medium \\
    \midrule
    \multicolumn{6}{c}{\textbf{CI Workflow Configuration}} \\
    \midrule
    Has docker build step
      & 2.70\%
      & 1.26\%
      & +114.9
      & ***
      & Negligible \\
    Uses cache
      & 37.83\%
      & 24.58\%
      & +53.9
      & ***
      & Small \\
    Matrix operating system count
      & 1.13
      & 1.24
      & $-$8.8
      & ***
      & Negligible \\
    Maximum matrix size
      & 20.43
      & 5.74
      & +255.8
      & ***
      & Negligible \\
    \midrule
    \multicolumn{6}{c}{\textbf{Build History and Context}} \\
    \midrule
    Last build duration, same CI workflow and branch (min)
      & 138.60
      & 17.92
      & +673.4
      & ***
      & \textbf{Large} \\
    Last build duration, same CI workflow (min)
      & 123.60
      & 17.89
      & +591.0
      & ***
      & \textbf{Large} \\
    Median duration of last 20 builds, same project (min)
      & 73.92
      & 15.92
      & +364.3
      & ***
      & \textbf{Large} \\
    Running builds of the same CI workflow
      & 1.77
      & 0.52
      & +240.6
      & ***
      & \textbf{Large} \\
    Duration variability of last 5 builds, same CI workflow and branch
      & 0.78
      & 0.39
      & +102.1
      & ***
      & Medium \\
    Triggered by push
      & 52.02\%
      & 25.87\%
      & +101.1
      & ***
      & Medium \\
    Failure rate of last 10 builds, same CI workflow and event
      & 0.29
      & 0.19
      & +51.1
      & ***
      & Small \\
    Finished earlier builds, same CI workflow and branch
      & 49.02
      & 49.66
      & $-$1.3
      & ***
      & Negligible \\
    Hours since last finished build, same project
      & 7.28
      & 8.90
      & $-$18.2
      & ***
      & Negligible \\
    Hours since last finished build, same CI workflow and branch
      & 66.27
      & 59.61
      & +11.2
      & ***
      & Negligible \\
    On default branch
      & 16.65\%
      & 17.79\%
      & $-$6.4
      & ***
      & Negligible \\
    \bottomrule
  \end{tabularx}
\caption*{\scriptsize Note:
\newline
(1)*** denotes $p < 0.001$, ** denotes $p < 0.01$, * denotes $p < 0.05$, and n.s.\ denotes $p \geq 0.05$.
\newline
(2) Effect sizes are interpreted using Cohen's $d$ as large ($|d| \geq 0.8$), medium ($0.5 \leq |d| < 0.8$), small ($0.2 \leq |d| < 0.5$), and negligible ($|d| < 0.2$).
\newline
(3) \textit{Has docker build step}, \textit{Uses cache}, \textit{Triggered by push}, and \textit{On default branch} are binary features, so the percentages reported in these rows indicate the proportion of builds in each group for which the feature value is 1~\citep{bennett2024developers}.
\newline
(4) Build history features use only builds that finished before the build was created. Builds without such earlier builds are left out of the corresponding row. Duration variability is the standard deviation of the logarithm of build duration.}
\end{table}

The CI workflows of the longest 18\% of builds also have more jobs than those of the remaining builds (9.49 vs.\ 2.66 DAG nodes on average).
Their largest jobs also have more steps (10.05 vs.\ 7.37 steps).
This result is consistent with prior work on build duration forecasting, which suggests that dependency structure is associated with incremental build duration~\citep{cao2017forecasting}.

However, CI workflow configuration features show a different pattern. \emph{Has docker build step}, \emph{Matrix operating system count}, and \emph{Maximum matrix size} differ significantly between the two groups, but all with negligible effect sizes. For \emph{Maximum matrix size}, the mean is much higher for the longest 18\% of builds (20.43 vs.\ 5.74), but the effect size is negligible.
\emph{Uses cache} is more common in the longest 18\% of builds (37.83\% vs.\ 24.58\%), but its effect size is small.
Taken together, the structure of a CI workflow is more consistently associated with the longest builds than its configuration.

\vspace{0.5em}
\noindent\textbf{Finding 3: Framework-specific CI workflows have the largest percentage of builds in the longest 18\%, and platform-specific CI workflows produce the most long builds.}
Table~\ref{tab:rq2-workflow-types} shows how often builds from each CI workflow category fall into the longest 18\% of builds.
\texttt{Framework-Specific} CI workflows have the largest percentage of builds in the longest 18\% (47.9\%), followed by \texttt{GPU \& CUDA} (36.0\%) and \texttt{Platform-Specific} CI workflows (32.7\%).
However, \texttt{Platform-Specific} CI workflows produce the most of these long builds. They have far more builds than the other two categories (71{,}485 builds), so they produce 48.1\% of the long builds.
In \texttt{Build \& Packaging} and \texttt{Testing} CI workflows, 14.6\% and 11.9\% of builds are in the longest 18\%. In \texttt{Code Lint} and \texttt{Documentation} CI workflows, only 1.6\% and 0.9\% are long builds\%.
\begin{table}[htbp]
  \scriptsize
  \centering
  \caption{Distribution of the longest 18\% of builds across CI workflow categories.}
  \label{tab:rq2-workflow-types}
  \begin{tabularx}{\linewidth}{%
      l
      >{\hsize=0.8\hsize\centering\arraybackslash}X
      >{\hsize=1.2\hsize\centering\arraybackslash}X
      >{\centering\arraybackslash}X
      >{\centering\arraybackslash}X
      >{\centering\arraybackslash}X}
    \toprule
    \textbf{Workflow Category}
      & \textbf{Builds}
      & \makecell[c]{\textbf{Rate in}\\\textbf{Longest 18\%}$^{*}$}
      & \makecell[c]{\textbf{Median}\\\textbf{Jobs Count}$^{\dagger}$}
      & \makecell[c]{\textbf{Matrix Jobs}\\\textbf{(\%)}$^{\ddagger}$}
      & \makecell[c]{\textbf{Parallel Jobs}\\\textbf{(\%)}$^{\S}$} \\
    \midrule
    Build \& Packaging
      & 31{,}609
      & 14.6\%
      & 4.0
      & 78.8\%
      & 93.0\% \\
    Platform-Specific
      & 71{,}485
      & 32.7\%
      & 4.0
      & 21.6\%
      & 61.4\% \\
    Testing
      & 98{,}264
      & 11.9\%
      & 3.0
      & 56.7\%
      & 49.2\% \\
    GPU \& CUDA
      & 10{,}637
      & 36.0\%
      & 3.0
      & 12.5\%
      & 82.3\% \\
    Documentation
      & 6{,}163
      & 0.9\%
      & 1.0
      & 0.0\%
      & 40.5\% \\
    Framework-Specific
      & 9{,}096
      & \textbf{47.9\%}
      & 6.0
      & 61.0\%
      & 78.2\% \\
    Code Lint
      & 42{,}676
      & 1.6\%
      & 1.0
      & 11.8\%
      & 26.1\% \\
    \bottomrule
  \end{tabularx}
  \caption*{\scriptsize Note:
  \newline
  $^{*}$ \textbf{Rate in Longest 18\%} is the percentage of builds in a CI workflow category that belong to the longest 18\% of builds.
  \newline
  $^{\dagger}$ \textbf{Median Jobs Count} is the median \emph{Jobs count} of builds in that CI workflow category.
  \newline
  $^{\ddagger}$ \textbf{Matrix Jobs (\%)} is the percentage of builds in that CI workflow category whose CI workflows use matrix jobs.
  \newline
  $^{\S}$ \textbf{Parallel Jobs (\%)} is the percentage of builds in that CI workflow category whose CI workflows contain parallel jobs.}
\end{table}

A \texttt{Framework-Specific} CI workflow checks that the project works correctly with a specific ML framework, such as PyTorch or TensorFlow.
As Table~\ref{tab:rq2-workflow-types} shows, \texttt{Framework-Specific} CI workflows have the highest median jobs count of all categories (6.0).
Their builds also include the cost of installing an ML framework and running the tests of the project with it.
Most \texttt{Framework-Specific} builds already use matrix jobs (61.0\%) and parallel jobs (78.2\%).

A \texttt{Platform-Specific} CI workflow builds and tests the project for a specific operating system or hardware architecture, such as Linux on ARM.
Only 21.6\% of \texttt{Platform-Specific} builds use matrix jobs, but 61.4\% use parallel jobs.
\texttt{Build \& Packaging} CI workflows use matrix jobs (78.8\%) and parallel jobs (93.0\%) more often than any other category, but only 14.6\% of their builds are long builds.
Therefore, the use of matrix jobs and parallel jobs does not separate the CI workflow categories with a large percentage of builds in the longest 18\% from those with a small percentage.

\begin{tcolorbox}[colback=gray!5!white, colframe=gray!75!black, title=Summary of RQ2, markedbreak]
  The longest 18\% of builds are most strongly associated with the duration of earlier builds of the same CI workflow.
  Their CI workflows also have more jobs and more dependencies between jobs.
  In contrast, CI workflow configuration features, such as caching and the size of the build matrix, show only small or negligible differences.
  Among CI workflow categories, \texttt{Framework-Specific} CI workflows have the largest percentage of long builds (47.9\%), and \texttt{Platform-Specific} CI workflows produce the most long builds (48.1\%).
  \end{tcolorbox}

\subsection{RQ3: Can the longest builds in ML projects be predicted accurately before execution?}
\paragraph{Motivation}
While the results of RQ1 and RQ2 show that a small subset of builds accounts for most of the total build time and that the longest builds are associated with CI workflow structure, they do not address a practical question for real-world development teams, \ie~can we predict which builds will belong to the top 18\% longest builds before they are triggered in ML projects?
Such predictive capability would enable proactive resource allocation, \ie~intelligent scheduling strategies for build jobs, and mitigate CI workflow bottlenecks in ML projects.
However, prior research on build duration prediction primarily relies on traditional metrics such as code churn and commit history~\citep{ghaleb2019empirical}. As RQ2 suggests, these metrics may miss CI workflow structure characteristics that are relevant to the longest builds in ML projects, especially CI workflow DAG complexity and the way jobs are organized within a CI workflow.
Therefore, we investigate whether features that describe the CI workflow structure, the recent builds of the same CI workflow, and the ML context of the project can improve prediction of the top 18\% longest builds over the traditional build feature baseline.
\paragraph{Approach}

To address RQ3, we formulate the prediction task as a binary classification problem where builds are labeled as belonging to the top 18\% longest builds (positive class) or the remaining builds (negative class) based on the threshold derived from RQ1 (a build duration of at least 39.20 minutes, which marks the top 18\% longest builds in our dataset).
We employ the set of 47 features described in Section~\ref{subsec:longbuilds:feature_engineering}, spanning run-level, CI workflow-level, and project-level features.
All features are constrained to information available at build submission time to prevent data leakage.
We train and evaluate four classification algorithms: Random Forest, XGBoost, LightGBM, and Logistic Regression.
Given the class imbalance, where the top 18\% longest builds represent only 18\% of all builds, we apply SMOTE~\citep{chawla2002smote,he2009learning} to balance the training data in each fold.
Because the models are meant to predict builds before they run, we evaluate them in chronological order rather than with random cross-validation, as recommended for software engineering data whose characteristics change over time~\citep{falessi2020need,bangash2020time}.
We sort all builds by their creation time and divide them into six blocks of equal size, similar to the online validation used for CI build failure prediction by~\citet{saidani2020predicting}.
In each fold, a model is trained on all blocks that precede a given block and tested on that block, so every test build was created after every training build.
We train each model on at least the first two blocks, \ie~one third of the builds, so the four folds test the third to the sixth block.
Because SMOTE and the learning algorithms are randomized, we repeat each fold with five random seeds~\citep{arcuri2011practical}, which gives 20 evaluations per model.
We evaluate model performance using six metrics: accuracy, precision, recall, F1-score, AUC-ROC, and AUC-PR.
We prioritize F1-score and AUC-ROC as primary indicators because they are more informative than accuracy under class imbalance~\citep{Saito2015}.

Because the global top 18\% threshold may not reflect the same meaning of longest builds in every project, we also conduct a robustness validation using project-level threshold definitions.
In this validation, we redefine the positive class (\ie~long builds) separately within each project based on that project's own build-duration distribution.
Specifically, for each project, we label the top 5\%, 10\%, 15\%, 18\%, 20\%, 25\%, and 50\% of builds in each project by duration as positive, and treat the remaining builds in that project as negative.
We then train and evaluate XGBoost within each project using the same set of 47 features, the same model settings, and the same SMOTE as in the main analysis.
Within each project, we follow the same chronological evaluation, dividing the builds of the project into six blocks, testing the third to the sixth block, and repeating each fold with five random seeds.
Because a single test block of a small project often contains no long build, we combine the predictions for the four test blocks of a project before computing the metrics.
We then summarize the metric distributions across projects.

\paragraph{Findings}
\begin{table}[htbp]
  \footnotesize
  \centering
  \caption{Prediction performance for the top 18\% longest builds.}
  \label{tab:rq3-model-performance-combined}
  \begin{tabularx}{\linewidth}{%
      l
      >{\centering\arraybackslash}X
      >{\centering\arraybackslash}X
      >{\centering\arraybackslash}X
      >{\centering\arraybackslash}X}
    \toprule
    \textbf{Metric}
      & \textbf{Random Forest}
      & \textbf{XGBoost}
      & \textbf{LightGBM}
      & \textbf{Logistic Reg.} \\
    \midrule
    \multicolumn{5}{c}{\textbf{Our full set of 47 features (including CI workflow, build history, and ML-specific features)}} \\
    \midrule
    F1
      & 0.7689 $\pm$ 0.0170
      & \textbf{0.7745 $\pm$ 0.0184}
      & 0.7734 $\pm$ 0.0185
      & 0.6862 $\pm$ 0.0265 \\
    AUC-ROC
      & 0.9540 $\pm$ 0.0091
      & \textbf{0.9554 $\pm$ 0.0079}
      & 0.9551 $\pm$ 0.0079
      & 0.9208 $\pm$ 0.0071 \\
    AUC-PR
      & 0.8112 $\pm$ 0.0133
      & \textbf{0.8217 $\pm$ 0.0146}
      & 0.8198 $\pm$ 0.0139
      & 0.7020 $\pm$ 0.0200 \\
    Precision
      & 0.6667 $\pm$ 0.0211
      & 0.7190 $\pm$ 0.0184
      & \textbf{0.7196 $\pm$ 0.0174}
      & 0.5779 $\pm$ 0.0295 \\
    Recall
      & \textbf{0.9085 $\pm$ 0.0127}
      & 0.8396 $\pm$ 0.0251
      & 0.8363 $\pm$ 0.0264
      & 0.8452 $\pm$ 0.0253 \\
    Accuracy
      & 0.8952 $\pm$ 0.0153
      & \textbf{0.9061 $\pm$ 0.0143}
      & 0.9060 $\pm$ 0.0141
      & 0.8524 $\pm$ 0.0148 \\
    \midrule
    \multicolumn{5}{c}{\textbf{Traditional build feature baseline proposed by~\citet{ghaleb2019empirical}}} \\
    \midrule
    F1
      & 0.6808 $\pm$ 0.0226
      & 0.6822 $\pm$ 0.0333
      & 0.6805 $\pm$ 0.0347
      & 0.5441 $\pm$ 0.0276 \\
    AUC-ROC
      & 0.9116 $\pm$ 0.0073
      & 0.9078 $\pm$ 0.0077
      & 0.9077 $\pm$ 0.0066
      & 0.8303 $\pm$ 0.0041 \\
    AUC-PR
      & 0.7074 $\pm$ 0.0147
      & 0.6977 $\pm$ 0.0176
      & 0.6958 $\pm$ 0.0205
      & 0.5436 $\pm$ 0.0460 \\
    Precision
      & 0.5867 $\pm$ 0.0172
      & 0.6144 $\pm$ 0.0294
      & 0.6208 $\pm$ 0.0305
      & 0.4082 $\pm$ 0.0238 \\
    Recall
      & 0.8115 $\pm$ 0.0393
      & 0.7686 $\pm$ 0.0535
      & 0.7554 $\pm$ 0.0596
      & 0.8164 $\pm$ 0.0379 \\
    Accuracy
      & 0.8545 $\pm$ 0.0171
      & 0.8640 $\pm$ 0.0115
      & 0.8654 $\pm$ 0.0114
      & 0.7384 $\pm$ 0.0270 \\
    \bottomrule
  \end{tabularx}
  \caption*{\scriptsize Note: Values are mean $\pm$ standard deviation over 20 evaluations (four chronological test blocks, each with five random seeds). Bold values indicate the best result per metric across all configurations.}
\end{table}

\begin{figure}[htbp]
    \centering
    \includegraphics[width=\linewidth]{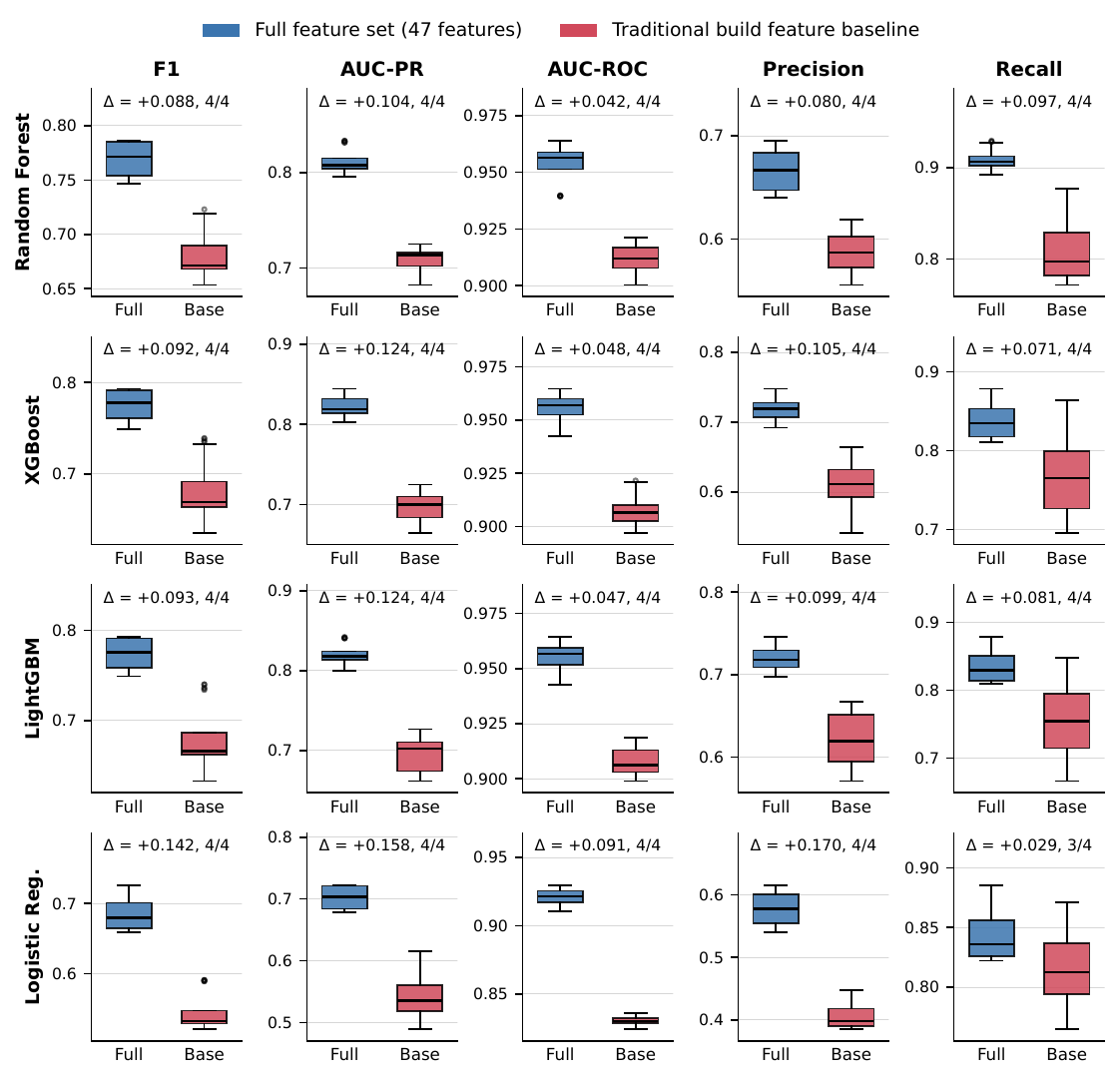}
    \caption[Full feature set vs.\ traditional build feature baseline]{Model performance comparison between the full feature set and the traditional build feature baseline.}
    \caption*{\footnotesize Note: Each box summarizes the 20 evaluations (four chronological test blocks, each with five random seeds) of one classifier and feature set. Each panel has its own y-axis. $\Delta$ is the mean difference (full feature set minus baseline) over the four test blocks, and the fraction is the number of blocks in which our proposed feature set is better.}
    \label{fig:rq3-comparison}
\end{figure}
\textbf{Finding 1: Our full set of 47 features outperforms the traditional build feature baseline proposed by \citet{ghaleb2019empirical} across all classifiers in our studied ML projects.}
Table~\ref{tab:rq3-model-performance-combined} and Figure~\ref{fig:rq3-comparison} compare the prediction results with two feature configurations.
With our full set of 47 features, all four classifiers achieve higher values than with the traditional build feature baseline on all six metrics.

XGBoost achieves the highest F1 with our full feature set, where F1 increases from 0.6822 to 0.7745 and AUC-PR increases from 0.6977 to 0.8217.
Random Forest and LightGBM improve by a similar amount, with F1 increasing from 0.6808 to 0.7689 and from 0.6805 to 0.7734, respectively.
The gain does not come at the cost of recall.
For XGBoost, precision increases from 0.6144 to 0.7190 and recall increases from 0.7686 to 0.8396, so the model finds more of the longest builds while raising fewer false alarms.

To check whether the improvement holds across time rather than only on average, we compare the two feature sets separately in each of the four chronological test blocks.
For all four classifiers and all six metrics, our full feature set performs better in each of the four test blocks, except for the recall of Logistic Regression, which is 0.0007 lower in one block.
For XGBoost, the F1 gain ranges from 0.056 to 0.136 across the four test blocks, and the AUC-ROC gain ranges from 0.036 to 0.061.
These results indicate that the improvement of our set of 47 features over the baseline is consistent across the four test blocks.

\textbf{Finding 2: The improvement is consistent across model families, indicating that the limitation of the traditional build feature baseline lies in the feature space rather than in the classifier.}
Logistic Regression, which is the same model used by \citet{ghaleb2019empirical}, improves from F1 of 0.5441 to 0.6862 and from AUC-PR of 0.5436 to 0.7020 when to our full set of 47 features. AUC-ROC also increases from 0.8303 to 0.9208. These three metrics are higher in all four test blocks.
The three tree-based classifiers (XGBoost, Random Forest, and LightGBM) show the same direction of change, with higher AUC-PR and AUC-ROC under our full set of 47 features. For example, the AUC-ROC of XGBoost increases from 0.9078 to 0.9554.

These results indicate that the longest builds in ML projects are better predicted when the features describe both how the CI workflow is organized and how the same workflow has behaved in recent builds, such as the duration of its last build on the same branch. Such information is absent from traditional build feature sets, which rely on code churn and commit history.

\begin{figure}[htbp]
    \centering
    \includegraphics[width=\linewidth]{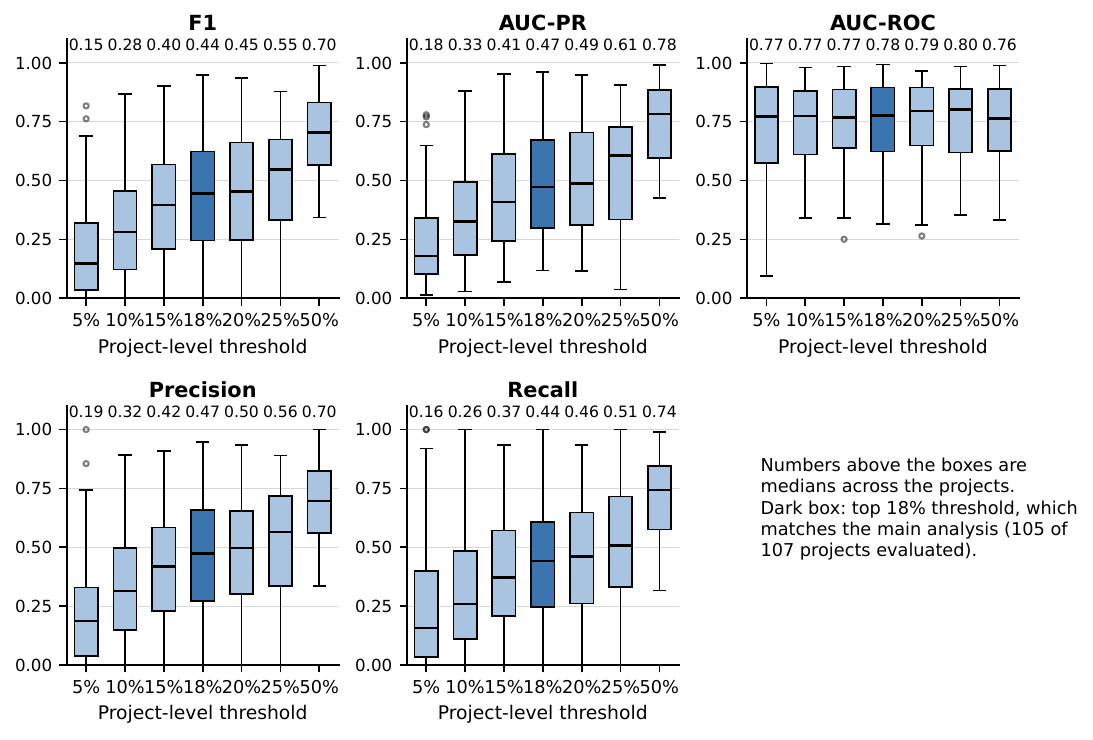}
    \caption{Prediction performance under project level percentile thresholds.}
    \caption*{\footnotesize Note: Each boxplot summarizes XGBoost performance across projects when long builds are defined using each project's own top 5\%, 10\%, 15\%, 18\%, 20\%, 25\%, or 50\% longest builds. Within each project, builds are evaluated in chronological order in the same way as in the main analysis, and each project value is the mean over five random seeds. A project is left out at a threshold when its test blocks contain no long build, which leaves 105 of the 107 projects at the top 18\% threshold. Numbers above the boxes are medians across projects.}
    \label{fig:rq3-project-threshold-comparison}
\end{figure}

\textbf{Finding 3: The model still separates relatively long builds when long builds are defined using project-level thresholds, although with lower performance than in the pooled analysis.}
Figure~\ref{fig:rq3-project-threshold-comparison} shows that XGBoost continues to separate relatively long builds from the other builds when each project uses its own percentile threshold instead of one global threshold.
Under the project-level top 18\% definition, the median F1 across projects is 0.4439, the median AUC-PR is 0.4714, and the median AUC-ROC is 0.7752.
These values are lower than those of the pooled model in Table~\ref{tab:rq3-model-performance-combined}. The pooled model learns from the builds of all projects, whereas each project-level model learns only from the earlier builds of a single project, which in small projects amount to a few dozen builds.
Performance increases as the threshold becomes less strict, with the median F1 rising from 0.1466 at the top 5\% threshold to 0.7028 at the top 50\% threshold, while the median AUC-ROC stays between 0.76 and 0.80 across all thresholds.
This result shows that the model's ability to rank longer builds above shorter ones is not specific to the single global top 18\% threshold used in the primary analysis.

\begin{tcolorbox}[colback=gray!5!white, colframe=gray!75!black, title=Summary of RQ3]
In our data, our full set of 47 features improves the prediction of the top 18\% longest builds over the traditional build feature baseline across all four classifiers when builds are evaluated in chronological order.
XGBoost performs best, with F1 increasing from 0.6822 to 0.7745, precision increasing from 0.6144 to 0.7190, and recall increasing from 0.7686 to 0.8396.
This suggests that long build prediction in ML projects benefits from features that capture how CI workflows are organized and how the same workflow has behaved in recent builds, rather than relying only on traditional build features.
When each project uses its own threshold, the model still ranks longer builds above shorter ones, although with lower performance than the model trained on all projects.
\end{tcolorbox}

\section{Discussion}
\label{sec:longbuilds:discussion}
This section explores the implications of our findings for developers and researchers, particularly in anticipating and managing the longest builds in ML projects.

\subsection{Implications for Developers}
\textbf{Developers should monitor the longest builds rather than average build duration.}
According to our findings, the long builds can account for most of the CI time in ML projects, and 39\% of them fail after the execution. Therefore, developers of ML projects can save much CI time by shortening the long builds or by organizing their jobs better. Teams can track how much of their CI time goes to long builds. When this time grows, teams can review the CI workflows that produce the long builds.

\textbf{Developers should review the structure of their CI workflows.}
In ML projects, the long builds run in CI workflows with more jobs and more dependencies between jobs. In contrast, caching, Docker builds, and the size of the build matrix differ little between the long builds and the other builds. When a CI workflow becomes slow, teams can first check whether each dependency between its jobs is needed. For example, a CI workflow may make all test jobs wait for a job that downloads pretrained model weights. However, the unit tests of data preprocessing code do not use these weights. These unit tests can start without waiting for the download. As another example, GPU test jobs may wait for all CPU test jobs to pass. This dependency keeps a failing commit from using expensive GPU runners. However, it also makes each build longer. Developers can decide which of the two matters more for each CI workflow. Teams can start with the CI workflows that test the project with a specific ML framework or build it for a specific platform. In our data, long builds are most frequent in framework-specific CI workflows, and platform-specific CI workflows produce the most long builds.

\textbf{Predictions of the longest builds can be used before a build starts.}
Our best model uses only features that are known before a build starts. It also predicts long builds well when it is trained on earlier builds and tested on later builds. For example, when a developer opens a pull request that changes the training code of an ML project, a CI service could warn the developer that the build is likely to take long. The developer can then work on other tasks instead of waiting for the build result.

\subsection{Implications for Researchers}
\textbf{Build duration studies should account for the type of ML project.}
Build duration differs mainly with how deeply a project uses ML, and differs less across ML domains. Projects that provide ML frameworks or infrastructure have longer builds than projects that only use ML libraries. Studies that treat all ML projects as one group may therefore miss this difference.

\textbf{Researchers should consider the history and structure of CI workflows as sources of features.}
Unlike the build features used in prior work, our features also describe the recent builds and the job structure of each CI workflow. With these features, our models predict long builds better than with the traditional build features~\citep{ghaleb2019empirical}. Future prediction models for CI could include features of recent builds and workflow structure. Future work could also test whether these features help predict build failures.

\section{Threats to Validity}
\label{sec:longbuilds:threats}
In this section, we discuss the threats to the validity of our study.

\subsection{Construct Validity}
Construct threats to validity concern whether our analyses measure what we claim to analyze.
Our study relies on build data collected from GitHub Actions and project metadata from GitHub repositories.
Inaccuracies in the collected data could influence our results.
To mitigate this threat, we apply filtering criteria to exclude incomplete or anomalous build records and validate our data extraction scripts through manual inspection of sampled builds.

Our classification of projects into ML usage depth and ML domain categories is performed through document analysis of repository descriptions, \texttt{README} files, and topic tags.
This manual classification process may introduce subjectivity.
To reduce this threat, three coders independently classify each project, as described in Section~\ref{approach:longbuilds:RQ1}.

Inter-rater agreement on the initial independent labels is high.
The remaining disagreements (10 of the 107 projects, 9.3\%) are resolved through discussion until consensus is reached.

Our main analyses focus on the longest 18\% of all builds across 107 ML projects.
This threshold is intended to study the builds that dominate total CI time, rather than to provide a universal definition of what counts as long in every individual project.
Different threshold choices may yield different results, so our conclusions should be interpreted with this scope in mind.

\subsection{Internal Validity}
Internal threats to validity concern the ability to draw conclusions about relationships between variables.
We study factors that are associated with the longest build durations using statistical tests and effect size measures.
We use the Mann--Whitney U test for continuous variables and the chi-square test for categorical variables, with effect sizes interpreted using Cohen's $d$.
However, statistical associations do not imply causation.
The observed relationships between CI workflow complexity and build duration may be confounded by other factors not captured in our analysis.

Our prediction models use features from CI workflow YAML files, earlier builds, and project metadata.
We base our 47 features on prior work and domain knowledge, but these features are not exhaustive.
Other factors such as code complexity, test suite characteristics, or hardware specifications may also influence build duration.
Our data also do not record runtime factors such as cache hits and runner availability. For the load on shared runners, we only count the running builds of the same workflow.
Our caching and parallelism features capture only how these mechanisms are configured in the workflow files, not how they behave at run time.
We therefore only claim that our features predict long builds better than the build features used in prior work.

The CI workflow DAG metrics are computed based on static analysis of YAML configuration files.
Runtime behavior, such as conditional job execution or dynamic matrix generation, may not be fully captured by our static analysis.
This could lead to discrepancies between the computed DAG structure and the actual execution graph.
In addition, we analyze one version of each CI workflow file, so all builds of the same CI workflow share the same workflow configuration features, even if the file changed during the study period.
We detect \emph{Has docker build step} and \emph{Uses cache} by keyword matching in the YAML text. Wrong matches may hide part of the difference that these two features show in RQ2.
Similarly, \emph{On default branch} treats only branches named \texttt{main} or \texttt{master} as the default branch, so builds on a default branch with another name, such as \texttt{develop}, are counted as builds on other branches.

\subsection{External Validity}
External threats concern the generalizability of our results.
Our study is based on 107 open-source ML projects on GitHub that use GitHub Actions.
Therefore, our conclusions may not generalize to ML projects using other CI platforms (e.g., Jenkins, GitLab CI, CircleCI) or to closed-source industrial ML projects.
Nevertheless, GitHub Actions is widely adopted in the open-source community, and our project selection procedure focuses on repositories drawn from previously curated ML project datasets, whose default branch was updated between January and July 2025, that contain GitHub Actions CI workflows triggered by code changes, and that have at least 100 CI runs and at least six months of CI history.
These criteria help us focus on actively maintained ML projects with sustained CI usage.

Our dataset is dominated by framework and infrastructure projects (72.0\%) and general-purpose ML projects (50.5\%).
Projects in underrepresented categories, such as reinforcement learning (1 project) or graph ML (2 projects), may not be adequately characterized.
Future work should investigate whether our findings hold for a more balanced distribution of ML usage depth and domain categories.

The workflow related features used in our prediction models are tailored to GitHub Actions workflow syntax.
Adapting these features to other CI platforms would require modifications to account for platform specific configuration formats and execution semantics.
Replication studies on other CI platforms are needed to assess the generalizability of our prediction approach.

\section{Related Work}
\label{sec:longbuilds:related}
This section reviews the prior work most directly related to the empirical contribution of this paper: characterizing and predicting the longest CI builds in ML projects.

\subsection{Long CI Builds and Build-Duration Prediction}

Prior work establishes build duration as an important CI performance concern.
\citet{laukkanen2015build} argue that long builds disrupt developer workflow because they delay the feedback that CI is intended to provide.
\citet{ghaleb2019empirical} model long build durations using project-relative thresholds and study how traditional build and project characteristics are associated with long builds.
Their feature set is especially relevant to this paper because it provides the traditional build duration baseline used in our RQ3 comparison.
\citet{ghaleb2022studying} further show that build duration and build breakages are related over time, suggesting that CI speed and CI reliability should not be treated as fully separate concerns.

Other studies predict or forecast build time from build histories.
\citet{bisong2017built} evaluate models for predicting Travis CI build time, while \citet{atchison2017time} and \citet{cao2017forecasting} study temporal patterns in CI build activity and duration.
These studies show that CI timing information can be modeled, but they mostly focus on general software projects and often treat a build as a single duration value.
They do not explain whether the longest builds in ML projects are concentrated in particular workflow categories, nor do they evaluate whether ML-specific workflow and project signals improve long build prediction.

\subsection{ML CI Workflows and Long-Build Evidence}

Recent studies show that ML projects have distinct CI/CD practices.
\citet{rzig2022characterizing} identify common CI tasks and failure modes in ML repositories.
\citet{bernardo2024machine} report that ML projects using GitHub Actions tend to have more complex workflows, more matrix configurations, and longer build durations than comparable non-ML projects.
\citet{bernardo2025continuous} provide developer evidence that model training, data preparation, dependency management, and infrastructure constraints often make CI in ML projects difficult to keep fast and stable.
\citet{houerbi2024empiricalarxiv} further show that ML CI/CD pipelines evolve through changes to build policies, dependencies, and workflow maintenance.

These studies motivate ML projects as a setting where CI duration deserves focused analysis.
However, these studies do not measure how much CI time the longest builds take. They also do not study which CI workflows produce these builds or whether these builds can be predicted before they run.
We address this gap by studying the longest 18\% of builds in 107 ML projects on GitHub Actions.

\subsection{CI Cost Reduction and Implications for ML Builds}

A related line of work reduces CI cost by selecting which tests or builds to run.
Regression test selection can shorten CI cycles by running only tests that are likely to be affected by a change~\citep{yu2018study,shi2019understanding,leong2019assessing}.
Build selection similarly aims to skip builds that are unlikely to reveal failures while preserving useful CI feedback~\citep{jin2022builds}.
These approaches are valuable, but they mainly target failure detection and test/build execution reduction.

Long builds in ML projects raise a complementary problem.
A build may pass and still delay feedback if it spends substantial time on dependency installation, data preparation, model-related execution, or artifact generation.
In such settings, skipping a short build and anticipating a long build have very different implications for developer waiting time and CI resource use.
Therefore, this paper complements CI cost-reduction research by focusing on whether the longest builds in ML projects can be identified and predicted, rather than only whether failing builds or useful tests can be selected.

\subsection{Position of This Study}

Prior work shows that build duration matters to developers and that it can be predicted. Other studies show that CI workflows in ML projects are more complex than those in many non-ML projects.
However, no prior study examines the longest builds in ML projects or predicts them before they run.
We address this gap in two steps. We first compare the CI workflows of the longest 18\% of builds with those of the other builds. We then test whether features from build history and workflow structure predict these builds better than the build features used in prior work.

\section{Conclusion}
\label{sec:longbuilds:conclusion}
In this paper, we study long builds in ML projects. We analyze 269{,}930 GitHub Actions builds from 107 open-source ML projects. The longest 18\% of builds take 80\% of the total build time, so we define these builds as long builds. Almost all long builds come from projects that provide ML frameworks or infrastructure. We then compare the CI workflows of long builds with those of the other builds. Long builds run in workflows with more jobs and more dependencies between jobs. The strongest association is with the duration of the previous build of the same workflow. Finally, we predict long builds before they run. Our 47 features predict long builds better than the build features used in prior work. We train four classifiers on earlier builds and test them on later builds. XGBoost performs best. With our features, it reaches an F1-score of 0.775, compared with 0.682 with the baseline features.

We report four main findings.

\begin{itemize}
    \item The longest 18\% of builds take 80.30\% of the total build time. Their failure rate is 39.09\%. In at least 75\% of the studied projects, the top 18\% longest builds account for more than half of the project's total build~time.

    \item
    Build duration depends on the type of ML project. Projects that provide ML frameworks or infrastructure have longer builds than projects that only use ML libraries. These projects produce 98.4\% of the long builds. General-purpose ML projects produce 63.6\% of the long builds, and scientific ML projects produce 24.1\%. The median build duration differs little between most ML domains.

    \item
    Long builds are most strongly associated with the duration of the previous build. Beyond build history, the structure of a CI workflow matters more than its configuration. Workflows with more jobs and more dependencies between jobs produce more long builds, while caching, Docker builds, and the size of the build matrix make little difference. The workflows that produce the most long builds run framework integration tests or build platform specific binaries.

    \item
    Long builds can be predicted in our studies of four classifiers, our proposed 47 features outperform the build features used in prior work. The best model XGBoost achieve the the F1-score at 0.775 by comparing with the baseline features at 0.682 .
\end{itemize}

Taken together, these findings show that the build features used in prior work do not fully explain long builds in ML projects.
Long builds are closely related to the recent builds and the job structure of each CI workflow.
For practitioners, reducing the dependencies between jobs may shorten builds more than changing the configuration of a workflow.

Future work could test whether these associations hold on other CI platforms and in industrial ML projects.
Another direction is to test whether reducing the dependencies between jobs actually shortens builds.
Finally, future work could add our prediction model to a CI service and study whether its warnings help developers.

\bibliographystyle{elsarticle-harv}
\bibliography{references}

\end{document}